\documentclass[12pt]{article}
\usepackage{ejorsj-s2}
\usepackage{amsmath,amssymb,amsfonts,amsthm}
\usepackage{graphicx}
\usepackage{color}
\usepackage{wrapfig}

\eqswitchon 
\newtheorem{problem}{Problem}[section]

\usepackage{mathrsfs}
\usepackage{algorithm}
\usepackage{algorithmic}
\usepackage{multirow}
\usepackage{hyperref}
\hypersetup{
    colorlinks   = true,
    citecolor    = blue,
    linkcolor=blue  
}
\usepackage{xstring}

\usepackage{color}
\usepackage{url}
\usepackage{here}

\usepackage{setspace}


\usepackage[multiple]{footmisc}
\usepackage{bigfoot}
\DeclareNewFootnote{AAffil}[arabic]

\date{\today}

\title{Long-Term Optimal Delivery Planning for Replacing the Liquefied Petroleum Gas Cylinder}

\author{
\begin{tabular}[h]{ccccc}  
Akihiro Yoshida & Haruki Sato & Shiori Uchiumi & Nariaki Tateiwa\thanks{ \ \ Present affiliation is NTT Software Innovation Center, NTT Corporation. } & Daisuke Kataoka \\ 
\multicolumn{5}{c}{\textit{Graduate School of Mathematics, Kyushu University}} \\
\multicolumn{5}{c}{Akira Tanaka} \\
\multicolumn{5}{c}{
\textit{
\begin{tabular}{c}
     Institute of Mathematics for Industry, Kyushu University\\
     National Institute of Information and Communications Technology
\end{tabular}
}
}
\\
\multicolumn{2}{c}{Nozomi Hata} & Yousuke Yatsushiro&Ayano Ide\thanks{\ \ Present affiliation is NTT Computer \& Data Science Laboratories, NTT Corporation.} & Hiroki Ishikura \\
\multicolumn{2}{c}{\textit{
\begin{tabular}{c}
     Institute of Mathematics for\\ Industry,
     Kyushu University
\end{tabular}
}
}
&
\multicolumn{3}{c}{\textit{Graduate School of Mathematics, Kyushu University}} \\
Shingo Egi & Miyu Fujii & Hiroki Kai  & \multicolumn{2}{c}{Katsuki Fujisawa} \\
\multicolumn{3}{c}{
\textit{
\begin{tabular}{c}
     Graduate School of Mathematics,\\
     Kyushu University
\end{tabular}
}
}
&
\multicolumn{2}{c}{
\textit{
\begin{tabular}{c}
     Institute of Mathematics for Industry,\\
     Kyushu University
\end{tabular}
}
}\\
\end{tabular}
}

\begin{document}

\maketitle
\begin{abstract}
In the daily operation of liquefied petroleum gas service, gas providers visit customers and replace cylinders if the gas is about to run out.
For a long time, frequent visits to customers were required because they could not determine the amount of remaining gas without a staff visit and observation.
To solve this problem, smart meters are started to be employed to acquire gas consumption more frequently without visiting customers.
In this study, we construct a system to optimize plans for cylinder replacement, and evaluate it with a large-scale field test.
We propose an algorithm to create a replacement plan with three steps: estimating the replacement date, acquiring the customer list for replacement, and determining the delivery route.
A more accurate estimation of the replacement date can be acquired with a smart meter, which is used for making a customer list for replacement.
The formulation for making a customer list enables the gas provider to replace cylinders some days before the date when the gas would run out.
It can suppress the concentration of replacements on certain days.
Large-scale verification experiments were performed with more than 1,000 customers in Chiba prefecture in Japan.
In the field test, the gas provider incorporated the system into its replacement operations.
Moreover, the replacement plans developed by the proposed system were compared with that by the gas provider.
Our system reduced the number of gas cylinders with gas shortage, the number of visits without replacement due to plenty of gas remaining, and the working duration per customer, which shows that our system benefits both gas providers and customers.
\end{abstract}





\section{Introduction}
To maintain liquefied gas service, a gas provider visits customers to verify the remaining amount of gas and replace cylinders if gas is about to run out.
We denote that the cylinder replacement problem (CRP) is the problem in making a plan to visit customers to replace gas cylinders, including the delivery route.
The CRP contains the following subproblems:
(1) estimating the replacement date to visit customers,
(2) making a customer list for replacement, and 
(3) obtaining the delivery route with a short working time.
Therefore, the CRP is so complex and requires many techniques to solve.

Owing to the difficulty of estimating the replacement date, smart meters have started to be installed for some customers to prevent a gas shortage.
They send the gas consumption to the gas provider more frequently than conventional meters, such as once a day.
For a long time before smart meters emerged, conventional meters were installed in customers' houses.
The gas provider is not able to determine the remaining amount of gas without a staff visit and observation.
Then, frequent visits by the gas provider are required to prevent a gas shortage, and many nonreplacement visits happen due to the plenty of remaining gas. 
In this situation, utilizing the smart meter's data instead of conventional meters can suppress the number of nonreplacement visits and gas shortages.
However, the gas provider cannot install the smart meter for all customers due to the installment costs.
Therefore, we still have to consider the problem of many nonreplacement visits for conventional meters.

\begin{figure}
    \centering
    \includegraphics[width=0.85\linewidth]{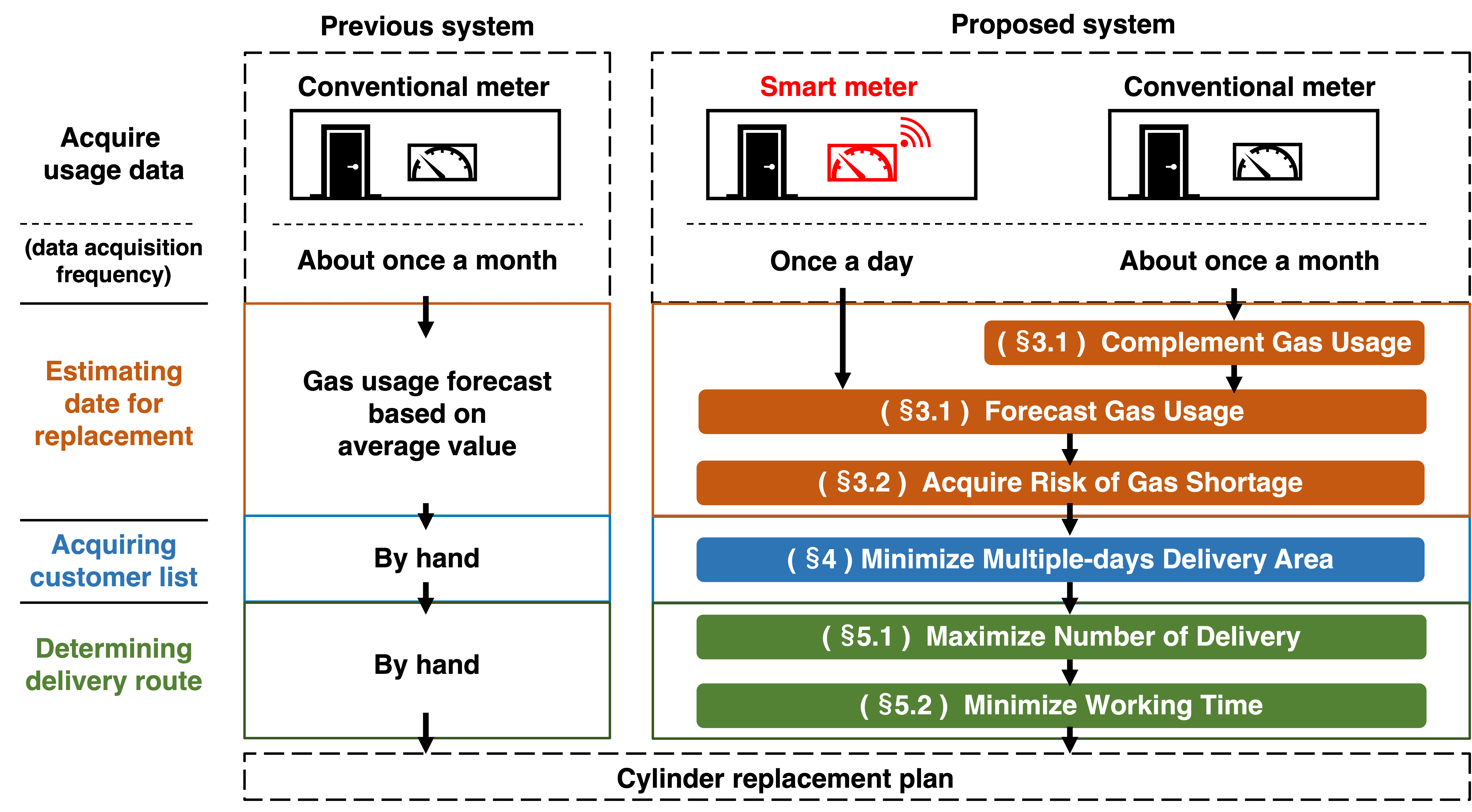}
    \caption{Proposed system for solving the gas cylinder replacement problem}
    \label{fig:overview}
\end{figure}

In this study, we construct a system to solve the CRP.
Our system enables us to make a gas cylinder plan without professional staff (see Figure~\ref{fig:overview}). 
Moreover, it suppresses the number of gas shortage and nonreplacement visits due to plenty of remaining gas, and decreases the working duration.
We can acquire a more accurate estimation of the replacement date with a smart meter, which is input for making a customer list for replacement.
Our formulation for making a customer list enables the gas provider to replace cylinders some days before the date about to the gas shortage.
Thus, the concentration of replacements on certain days can be avoided.

We briefly explain the proposed algorithm to solve the CRP with three steps, and the overall flow and the comparison with the previous operation are shown in Figure~\ref{fig:overview}. 
First, we estimate the cylinder's replacement date. (Section~\ref{sec:demand_forecast}).
The risk of gas shortage per day is calculated based on the current remaining gas and multiple days of forecast gas usage.
Beforehand, we extrapolate the conventional meter's gas usage to estimate the current remaining gas because the gas provider does not observe the current remaining gas for the conventional meter.
After estimating the replacement date, a customer list for replacement is acquired (Section~\ref{sec:delivery_list}).
When we greedy pick up all customers for the predicted replacement date, there can be too many customers to visit or a long working duration.
To prevent this situation, the multiple-day customer list is considered simultaneously.
Then, we regard the problem as obtaining the smallest rectangle that covers the customers to need replacements, and we formulate the problem and solve it as a mixed-integer optimization problem.
Finally, the order of visits to replacement is determined (Section~\ref{sec:delivery_route}).
The replacement should be conducted as much as possible to prevent the many customers close to a gas shortage at the same time.
We also formulate and solve the problem as mixed-integer optimization problems for obtaining the delivery route with the shortest working duration.

To verify whether the proposed system is useful, two experiments were conducted with more than 1,000 customers, of which approximately one-third did not have smart meters.
In a field test, a gas provider followed the output of our system and succeeded in reducing the nonreplacement visits and working duration (Section~\ref{sec:field_test}).
Furthermore, we performed a verification experiment by comparing the cylinder replacement plan made by the proposed system and that developed by the gas provider (Section~\ref{sec:exp_dldr}).
The proposed system reduced the number of out-of-gas cylinders, number of visits without replacement due to the plenty of gas remaining, and working duration, which shows that it benefits both gas providers and customers.
\section{Related Work\label{sec:related}}
In this study, the cylinder replacement problem, which is a specific case of the gas secondary distribution problem, is addressed.
A secondary distribution is the process of transporting refined oil from oil depots to petrol stations or end-users.
When we acquire the customer list for the replacement and delivery route, the output of the machine learning such as demand forecast is often utilized.
In this case, it is necessary to consider the uncertainty of machine learning to prevent an unfavorable situation, such as items out of stock.
In this section, we explain the related work about the secondary gas distribution problem and delivery planning with considering demand forecast's uncertainty.

Most studies on secondary gas distribution problems have addressed the gas replenishment problem, whereas only a few have explored the gas cylinder replacement problem.
\cite{carotenuto2015periodic} decomposed the final distribution of fuel oil into a weekly replenishment plan for each station and arranged petrol station visiting sequences (vehicle routes) for each day of the week. 
\cite{cornillier2012heuristics} proposed a multi-depot petrol station replenishment problem with time windows. 
Multiperiod optimization of the gas is an important property to consider in a sequential decision in \cite{cornillier2008heuristic, hanczar2012fuel, charusakwong2016optimization}.
Like the gas replenishment problem, the research for making customer lists and delivery route for other applications has also been widely explored.
Note that these problems belong to the inventory routing problem.
\cite{hernandez2017heuristics} also studied a tactical delivery plan for large retailers selling and delivering large and heavy items to customers, and \cite{lagana2021dynamic} modeled a general dynamic multi-period routing problem for postal and courier companies. 
A similar problem was addressed by~\cite{ulmer2018value}, who examined a dynamic multi-period vehicle routing problem with stochastic service requests. 
In this article, the multiperiod plan described in Section~\ref{sec:delivery_list} is also considered.

Because many delivery plans are made based on the output of the demand forecast, the other important property to consider in optimal delivery planning is the demand forecast uncertainty. 
\cite{singh2019multi} used a multiscenario mixed-integer programming model to describe the problem of the transportation process of oil products, considering a stochastic hub disruption and uncertain demands.
In addition,~\cite{li2020two} presented a two-stage stochastic programming model that determines the replenishment quantity of each petrol station.
Moreover,~\cite{bertazzi2015managing} regarded the demands of retailers as discrete random variables to consider the stochastic inventory routing problem with transportation procurement.
\cite{nikzad2019two} proposed mixed-integer programming models for medical drug distribution problems under demand uncertainty, such as drug storage. 
In this study, the demand forecast's uncertainty, which is described in Section~\ref{sec:risk_func}, is also considered.

\section{Estimating Replacement Date for Customer\label{sec:demand_forecast}}
In this section, we describe the estimation of the replacement date for customers through three steps.
As a result of the estimation, the customers are categorized into three groups per day, high-risk customers, moderate-risk customers, and low-risk customers, based on the emergency of the gas shortage. The categorization is utilized for acquiring the customer list (see Figure~\ref{fig:overview}).
First, the gas usage for conventional meters is extrapolated, as described in Section~\ref{sec:demand_forecast_sub}.
In particular, daily gas consumption data from smart meters are utilized to extrapolate gas consumption data for conventional meters.
The second step is to forecast gas usage.
Well-known machine-learning models are employed for customers with smart meters, such as support vector regression and random forest regression.
For conventional meters, daily gas usage is forecasted using a proposed algorithm based on the $k$-nearest-neighbor algorithm, which is also described in Section~\ref{sec:demand_forecast_sub}.
Finally, we estimate the emergency of replacement per day.
The customers are categorized into three groups by considering the forecast gas usage and the demand forecast uncertainty, which is described in Section~\ref{sec:risk_func}.

\subsection{Forecasting and Complementing of Gas Usage for Conventional Meters\label{sec:demand_forecast_sub}}
\begin{figure}
    \centering
    \includegraphics[width=0.8\linewidth]{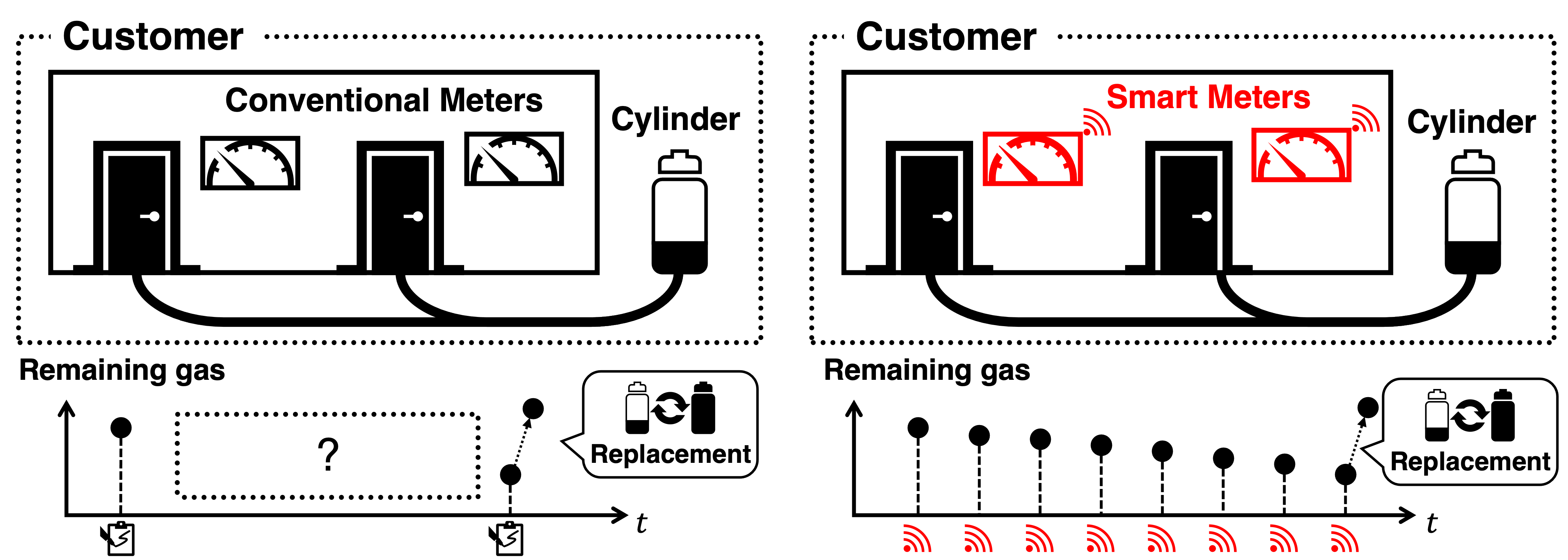}
    \caption{Relationship among meter, cylinder, and customer. Note that the remainder of gas is observed approximately once a month for conventional meters while per day for smart meters.}
    \label{fig:customer}
\end{figure}

We propose a complement and forecast method for daily gas usage based on data acquisition from a one-month period.
First, the set of customers and meters are defined as follows.
Let $\mathcal{M}, \mathcal{M}_s, \mathcal{M}_c$ be a set of meters, smart meters, and conventional meters, respectively. 
Because every meter is either smart meters or conventional meters, $\mathcal{M} = \mathcal{M}_s \sqcup \mathcal{M}_c \ (\mathcal{M}_s \cap \mathcal{M}_c = \emptyset)$ is satisfied.
Note that $M_c \in \mathcal{M}_c$ needs extrapolation to predict the current remaining gas, whereas $M_s \in \mathcal{M}_s$ does not.
Some households share the same gas cylinders, and the collection of them is considered a customer.
Let $\mathcal{C}$ be the set of customers, and we denote the set of meters $m(C) \subset \mathcal{M}$ corresponds to the customer $C$.
Because each household has a meter to record the gas usage, the total gas usage for the customer is the summation of the gas usage of the meters.
The relationship among meters, cylinders, and customers is shown in Figure~\ref{fig:customer}.
The gas usage is forecasted per meter $M \in \mathcal{M}$,  whereas the replacement is performed per customer $C \in \mathcal{C}$.

The daily gas usage of the meter $M$ for date $D$ is denoted as ${\mu}_{D}^{(M)}$.
For a smart meter $M_s$, ${\mu}_{D}^{(M_s)}$ can be acquired by the smart meter. 
However, for a conventional meter $M_c$, the gas usage can be observed approximately once a month.
Therefore, ${\mu}_{D}^{(M_c)}$ is acquired through interpolation based on the two successive gas usage observations. 
Moreover, the forecast gas usage of the meter $M$ for the date $D^{\prime}$ is denoted as $\widehat{\mu}_{D^{\prime}}^{(M)}$.

Daily gas consumption data from smart meters are used to extrapolate and forecast gas consumption data for conventional meters.
We propose how to extrapolate and forecast daily usage for a conventional meter whose basic concepts are divided into two steps.
First, for every conventional meter, $k$ smart meters are extracted such that the usage tendency is similar to that of the conventional target meter.
Second, the extrapolated and forecast value are obtained by the weighted average of their usages.

For a conventional meter $M_c$, the last observed date is denoted as $D_{M_c}$.
A function ${\rm Sim}:{\mathcal M}_c \times \mathcal{M}_s \rightarrow \mathbb{R}$ is defined to measure the similarity of gas usage tendency between a conventional meter $M_c \in \mathcal{M}_c$ and a smart meter $M_s \in \mathcal{M}_s$ as follows, 
\begin{equation}
    {\rm Sim}(M_c,M_s):=
    \frac{1}{\sum_{i=0}^{n_s} \left({\mu}_{D_{M_c}-i}^{(M_c)} - {\mu}_{D_{M_c}-i}^{(M_s)} \right)^2}
\end{equation}
where $n_s$ is the hyperparameter and $D_{M_c} - i$ indicates the date $i$ days before date $D_{M_c}$.
Based on the gas usage similarity, the \emph{k}-nearest neighbor meters are extracted for every conventional meter $M_c$ and we denote it as $NN(M_c,k)$.

The extrapolated gas usage at $D$ for $M_c$ is obtained by calculating the weighted average of the observed gas usage of $NN(M_c, k)$ as follows:
\begin{eqnarray}
    &{\mu}_{D}^{(M_c)} = \sum_{M_s\in NN(M_c,k)}{\rm Sim}^*{(M_c,M_s)} \  {\mu}_{D}^{(M_s)} \\ 
    &{\rm where} \ {\rm Sim}^*{(M_c,M_s)} = \frac{{\rm Sim}{(M_c,M_s)}} {\sum_{M_s^{\prime} \in NN(M_c,k)} {{\rm Sim}{(M_c,M_s^{\prime})}}}, \ M_s \in NN(M_c,k)  .
\end{eqnarray}
The gas usage forecast at $D^{\prime}$ for $M_c$ is also obtained by calculating the weighted average of the forecast gas usage of $NN(M_c, k)$ as follows:
\begin{equation}
    \widehat{\mu}_{D^{\prime}}^{(M_c)} = \sum_{M_s\in NN(M_c,k)}{\rm Sim}^*{(M_c,M_s)} \  \widehat{\mu}_{D^{\prime}}^{(M_s)}
\end{equation}
where $\widehat{\mu}_{D^{\prime}}^{(M_s)} (M_s \in \mathcal{M}_s)$ can be obtained by the well-known machine-learning methods, such as support vector regression or random forest regression.

Although the concepts are close to the well-known $k$-nearest neighbor algorithm for regression, to the best of our knowledge, this application has not been previously explored.
\subsection{Estimating Risk of Gas Shortage by Considering Gas Usage Forecast's Uncertainty\label{sec:risk_func}}

Based on the gas usage forecast, we determine whether each customer should be visited or not by obtaining the value of the proposed risk function.
The risk function is defined by considering the gas usage forecast and the forecast uncertainty to quantify the emergency of a cylinder replacement.
Based on the value of the risk function, high- and moderate-risk customers were extracted as candidates for replacement, which is the input for the problem of making a customer list for replacement.

The definition of the proposed risk function is introduced mathematically.
The forecast error of each meter per day is assumed to independently follows a normal distribution.
Obviously, the daily gas consumption of one meter does not influence that of the other meters.
First, the gas usage forecast and unbiased variances are summed up for some meters belonging one customer per day.
In sequence, the cumulative summation is calculated to obtain the cumulative gas usage and unbiased variances for a customer.
Based on this information, the risk function is defined as the probability of falling below a specific gas rate. 

Letting $D_0$ be the current date, we select one customer $C \in \mathcal{C}$.
The risk of gas shortage after $n_f$ dates is considered.
$\widehat{\mu}_{{D_0}+i}^{(M)}$ and $\widehat{\sigma^2_{i}}^{(M)}$ are indicated the forecast daily gas consumption for the $i$-th day after $D_0$ and an unbiased variance for the forecast gas consumption for $i$ dates, respectively.
Then, $\mathcal{N}\left(\widehat{\mu}_{{ D_0}+i}^{(M)}, \widehat{\sigma^2_{i}}^{(M)} \right) (M \in m(C)) $ is the probability distribution function for forecasting the gas consumption on the $i$-th day after $D_0$ for meter $M$~($i \in \{0,1,\cdots,n_f-1\}$), where $\mathcal{N}(\mu, \sigma^2)$ is the normal distribution with the parameters $\mu$ and $\sigma$ indicating the average and the standard deviation, respectively.
It is derived from the assumption that the error of the gas usage forecast follows a normal distribution.
Based on the assumption that the error in the gas usage forecast of each meter per day independently follows a normal distribution, the reproductive property of the normal distribution can be applied.
Therefore, the function representing the amount of gas consumption distribution $p_{D_0, n_f}^{(C)}$ for $n_f$ days from date $D_0$ can then be written as follows:
\begin{equation}
    X_{D_0, n_f}^{(C)} \sim p_{D_0, n_f}^{(C)} = \mathcal{N}\left(\sum_{i=0}^{n_f - 1}\sum_{M \in m(C)} \widehat{\mu}_{D_0+i}^{(M)}, \sum_{i=0}^{n_f -1}\sum_{M \in m(C)} \widehat{\sigma^2_{i}}^{(M)}\right)
\end{equation}
where $X_{D_0, n_f}^{(C)}$ is the random variable indicating the gas usage of customer $C$ from date $D_0$ to $n_f$-th day after $D_0$.

Let $s^{(C)}_{D_0}, \varepsilon_\alpha^{(C)}$ be the remaining gas of customer $C$ at date $D_0$ and the amount of gas of customer $C$ when the remaining gas rate is equal to $\alpha$.
$s^{(C)}_{D_0}$ can be acquired by utilizing $\mu_{D_0 - i}^{(M)} \ \left(\forall M \in m(C), \forall i \ {\rm s.t.} \ D_0-i \ge D_{M}\right)$ and the latest cylinder replacement date.
The function $\hat r_{\alpha, D_0}^{(C)} : \mathbb{N} \to [0,1]$ can be described, which returns the probability that the remaining gas rate of customer $C$ is less than $\alpha$ on given days after date $D_0$, where $P$ is the probability distribution function.
\begin{align}
\hat r_{\alpha, D_0}^{(C)}(n_f) :=& \ P\left(X_{D_0, n_f}^{(C)} \ge s^{(C)}_{D_0} - \varepsilon_\alpha^{(C)}\right) 
\end{align}

Replacement for the customer cannot be conducted every day owing to the staff vacation or the customer's availability.
Therefore, it is necessary to consider whether or not to replace the customer's cylinder such that a gas shortage does not occur during the nonavailability dates.
Therefore, the risk on $n_f$-th day after $D_0$ is assessed based on the remaining gas on the first available date after $n_f$-th day after $D_0$. 
Formally, if the $I_{\rm avail}^{(C)} :  \mathcal{D} \to \{0,1\}$ is prepared as the indicator function representing the availability of customer $C$ including staff's availability, where $\mathcal{D}$ is the set of dates, the risk function $r_{\alpha, D_0}^{(C)} :  \mathbb{N} \to [0,1]$ representing the emergency of replacement for the customer $C$ on given days after date $D_0$ is defined as follows:
\begin{equation}
    r_{\alpha,D_0}^{(C)}(n_f) := \hat r_{\alpha, D_0}^{(C)}(n_f^*) \ \left(n_f^* = \underset{}{\rm min}\ \{ i \in \mathbb{Z} \mid i \ge n_f,  I_{\rm avail}^{(C)}(D_0+i+1)=1 \} \right)
\end{equation}

In this case, two thresholds are prepared for the remaining gas rate: $\alpha_{\textrm{high}}$ and $\alpha_{\textrm{mdr}}$ for categorizing the customers into three groups.
When we prepare the thresholds $q_{\textrm{high}}, q_{\textrm{mdr}}$, the high- and moderate-risk customers on $i$ days after $D_0$ can be written as follows:

\begin{itemize}
    \item High-risk customers $\mathcal{C}_{\textrm{high}, D_0}(i) := \{C \in \mathcal{C} \mid  r_{\alpha_{\textrm{high}}, D_0}^{(C)}(i) \ge q_{\textrm{high}}  \}$
    \item Moderate-risk customers $\mathcal{C}_{\textrm{mdr},D_0}(i) := \{C \in \mathcal{C} \setminus \mathcal{C}_{\textrm{high}, D_0}(i) \mid  r_{\alpha_{\textrm{mdr}}, D_0}^{(C)}(i) \ge q_{\textrm{mdr}} \}$
\end{itemize}

Customers except for high-risk customers and moderate-risk customers are categorized as low-risk customers.
Its mathematical notation is omitted because they cannot be utilized later in the system.
We simply write high- and moderate-risk customers as $\mathcal{C}_{\textrm{high}}(i)$ and $\mathcal{C}_{\textrm{mdr}}(i)$, respectively, when it is unnecessary to take care of $D_0$.
When the customer list for replacement is acquired using the above definition, moderate-risk customers are occasionally relatively small in number.
To increase the number of moderate-risk customers, we set a parameter $D_{lbd}$. 
It ensures that the customer is categorized as a moderate-risk customer  $D_{lbd}$ days before becoming a high-risk customer.
\section{Acquiring Customer Lists for Multiple Dates for Replacement by Minimizing  Delivery Area\label{sec:delivery_list}}

In this section, the acquisition of customer lists for replacement is discussed through categorizing customers as high risk, moderate risk, or low risk.
We formulate the problem of acquiring the customer list as acquiring the smallest rectangle that covers the customer lists.
The formulation enables the gas provider to replace cylinders some days before the date about to the gas shortage, thereby suppressing the concentration of replacements on certain days.
Therefore, multiple-day customer lists are obtained simultaneously.

\begin{wrapfigure}[13]{r}[3mm]{0.45\linewidth}
    \centering
    \includegraphics[width=0.8\linewidth]{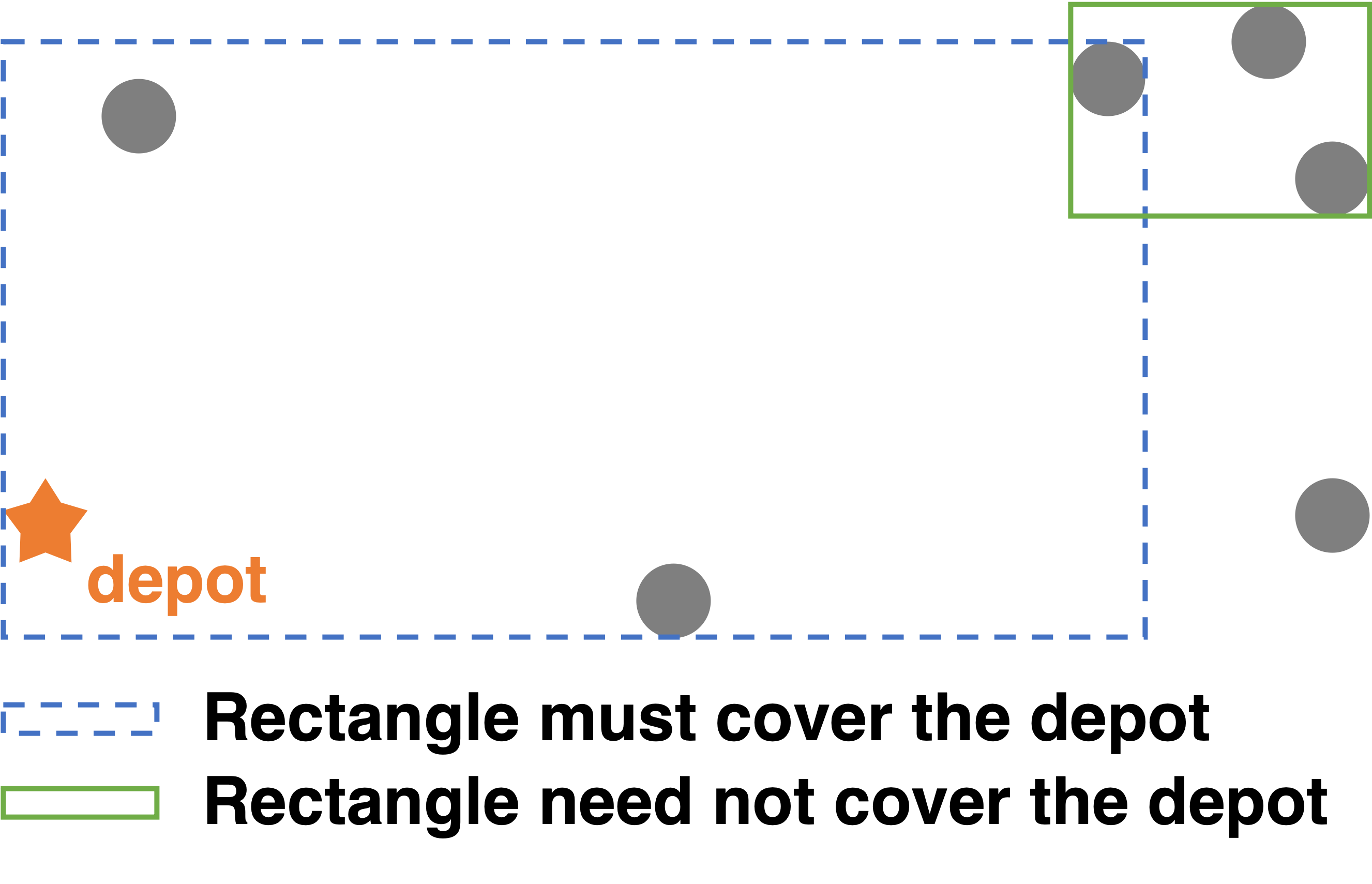}
    \caption{Illustration of the influence whether the depot is included or not in the output}
    \label{fig:include_depot}
\end{wrapfigure}

We call a {\it trip} the process that the truck leaves the delivery center, visit customers to replace the gas cylinders, and returns to the delivery center.
When we denote the planning horizon as $D_{ph}$, we consider the problem for the dates $ \{D_0, D_0+1 ,\cdots , D_0 + D_{ph}\}$.
There are two types of cylinders, large-sized cylinders and small-sized cylinders, and each customer has one type of cylinder.
Note that the formulation can be easily extended to situations where there are more types of cylinders or customers have several types of cylinders.
Our formulation is based on the following concepts:

\begin{enumerate}
    \item Bounding-box based model\\
    We formulate the problem to acquire the delivery list as obtaining the small rectangle that covers the customers needing replacement cylinders.
    The rectangle is prepared per trip for each date.
    In particular, the formulation does not ensure that the rectangle covers the depot, unlike in existing approaches~\cite{baller2019dynamic}.
    Figure~\ref{fig:include_depot} illustrates the difference in the customer list derived from the formulation regardless of whether the depot is included.
    When a customer far from the depot must be visited, the rectangle becomes large with the previous approach, and the customer requiring delivery must not be gathered. 
    A numerical experiment about these differences is discussed in Appendix~\ref{sec:exp_ablation}.
    \item Multi-period optimization model\\
    When the delivery route per day is obtained sequentially, it may harm the future delivery area by resulting in a large delivery area.
    To prevent such occurrences, we simultaneously acquire a multi-day delivery list.
\end{enumerate}

The formulation is shown in Problem~\ref{prob:multiBB}.
We attempt to minimize the maximum value of the summation of the length of edges in a rectangle.
Constraints (1)--(4) ensure that the rectangle covers the customers that have been shortlisted for cylinder's replacement.
Constraint (5) enables us to deliver consistently to high-risk customers.
Without constraint (5), the output rectangle has no incentive to cover moderate-risk customers.
Therefore, it is necessary to determine the lower bound of the number of deliveries $LB$.
Constraint (6) ensures at least one visit to a high-risk customer during the target days.
Every customer can belong to at most one trip each day, which is ensured by constraint (7).
Constraint (8) ensures truck availability in terms of maximum weight loading.
Constraints (9) and (10) are prepared to satisfy the truck availability in terms of space.
Let $N_{\textrm{Large}}$ be the maximum number of large-sized cylinders being carried with a single truck, and the constraint (9) describes the maximum number of large-sized cylinders.
Let $\gamma_{l}$ be the maximum number of small-sized cylinders given the number of large-sized cylinders $l \ (l = 0, 1, \cdots, N_{\textrm{Large}})$ being loaded.
There is an almost linear relationship between $l$ and $\gamma_{l}$ based on the observation.
Then, the linear function $a l + b$ is prepared with parameters $a$ and $b$ such that the following inequality is satisfied $a  l + b \le \gamma_l \ (l = 0, 1, \cdots, N_{\textrm{Large}})$.
To maintain the search space as much as possible, the gap between $al + b$ and $\gamma_l$ should be as small as possible among all candidates, approximately $a$ and $b$.
The optimal parameters among them are denoted as $a^*$ and $b^*$.
Finally, a constraint (10) is added for the number of small-sized cylinders with linear inequality.
When the problem is infeasible, the situation is derived from constraint (5) because the numbers of high-risk and moderate-risk customers are fewer than the lower bound.
Under this situation, we reduce the lower bound by 1, and solve the problem repeatedly until the solution is obtained.

\begin{enumerate}
\item Variables
    \begin{itemize}
        \item $y^{(C)}_{i, j}=
        \left\{
            \begin{array}{ll}
                1 & \text{Customer} \ C \ \text{is selected in the delivery list with ${j}$-th trip on after $i$ days} \\
                0 & \text{otherwise}
            \end{array}
        \right.$
        \item $v^{lb}_{i,j}, v^{ub}_{i,j}, h^{lb}_{i,j}, h^{ub}_{i,j} \in \mathbb{R}$: Variables representing a rectangle covering the delivery area for $j$-th trip after $i$ days.
    \end{itemize}
\item Constants
    \begin{itemize}
        \item $a^*, b^* \in \mathbb{R}$: Hyperparameters representing the available truck capacity
        \item $W > 0$: Hyperparameters representing the maximum loading weight of the truck
        \item $\mathcal{C_{\rm Large}} \subset \mathcal{C}$: Set of customers where large-sized cylinders are installed
        \item $N_{\rm Large} \ge 0$: Maximum number of large-sized cylinders being carried with a single truck 
        \item $\mathcal{C}_{\textrm{high}}(i) $: High-risk customers after $i$ days
        \item $\mathcal{C}_{\textrm{mdr}}(i) $: Moderate-risk customers after $i$ days
        \item Location for customer $C$: $(\textrm{lon}^{(C)}, \textrm{lat}^{(C)}) \in \mathbb{R}^2_{\ge 0}$
        \item Number/Weight of cylinders for customer $C$ : $\textrm{cn}^{(C)}, \textrm{cw}^{(C)}$
        \item Lower bound of number of visiting customer : $LB$
        \item Planning horizon (number of dates) : $D_{ph}$
        \item $\mathcal{\theta}_{\textrm{high}}^{(C)} := \min\left\{i \mid C \in \mathcal{C}_{\textrm{high}}(i)\right\}$
        \item $\mathcal{\theta}_{\textrm{mdr}}^{(C)} := \min\left\{i \mid C \in \mathcal{C}_{\textrm{mdr}}(i) \sqcup \mathcal{C}_{\textrm{high}}(i) \right\}$
    \end{itemize}
\end{enumerate}

\begin{problem}
\label{prob:multiBB}
\rm
Obtaining the smallest rectangle that covers the customers for replacement to acquire multiple-days' customer list

\begin{align}
& \underset{y, v, w, L}{\rm minimize} && \underset{i, j}{\rm max}\left\{\left(v^{ub}_{i, j}-v^{lb}_{i, i}\right) + \left(h^{ub}_{i, j}-h^{lb}_{i, j}\right)\right\} \notag
\\
& \rm{subject \ to} && \textrm{lon}^{(C)} y^{(C)}_{i, j} \le h^{ub}_{i, j} \ \tag*{$(\forall i, \forall j, \forall C \in \mathcal{C}_{\textrm{mdr}}(i) \sqcup \mathcal{C}_{\textrm{high}}(i))$ \ (1)}
\\
& && {\rm lon}^{(C)} y^{(C)}_{i, j} + \underset{C^{\prime} \in \mathcal{C}}{\rm max}\ {\rm lon}^{(C^{\prime})} \left(1-y^{(C)}_{i, j}\right)  \ge h^{lb}_{i, j} \ \tag*{$(\forall i, \forall j, \forall C \in \mathcal{C}_{\textrm{mdr}}(i) \sqcup \mathcal{C}_{\textrm{high}}(i))$ \ (2)} 
\\
& && {\rm lat}^{(C)} {y^{(C)}_{i, j}} \le v^{ub}_{i, j} \  \tag*{$(\forall i, \forall j, \forall C \in \mathcal{C}_{\textrm{mdr}}(i) \sqcup \mathcal{C}_{{\textrm{high}}}(i))$ \ (3)}
\\
& && {\rm lat}^{(C)} y^{(C)}_{i, j} + \underset{C^{\prime} \in \mathcal{C}}{\rm max}  \ {\textrm{lat}^{(C^{\prime})}} \left(1-y^{(C)}_{i, j}\right)  \ge v^{lb}_{i, j} \ \tag*{$(\forall i, \forall j, \forall C \in \mathcal{C}_{\textrm{mdr}}(i) \sqcup \mathcal{C}_{\textrm{high}}(i))$ \ (4)}
\\
& && \sum_{C \in \mathcal{C}_{\textrm{mdr}}(i) \sqcup \mathcal{C}_{\textrm{high}}(i)} y^{(C)}_{i, j} \ge LB \ \tag*{$(\forall i, \forall j)$ \ (5)} 
\\
& && \sum_{i = \theta_{\textrm{mdr}}^{(C)}}^{\theta_\textrm{high}^{(C)}} \sum_{j} {y^{(C)}_{i, j}} \ge 1 \ \tag*{$\left(\forall C \in \mathcal{C} {\rm ~s.t.~}  \theta_{\textrm{high}}^{(C)} \le D_{ph} \right)$ \ (6)}
\\
& && \sum_{j} y^{(C)}_{i, j} \le 1 \ \tag*{$(\forall i, \forall C \in \mathcal{C}_{\textrm{mdr}}(i) \sqcup \mathcal{C}_{\textrm{high}}(i))$ \ (7)}
\\
& && \left(\sum_{\mathcal{C}_{\textrm{mdr}}(i) \sqcup \mathcal{C}_{\textrm{high}}(i)} {\rm cw}^{(C)} {\rm cn}^{(C)}y^{(C)}_{i,j}\right) \le W \  \tag*{$(\forall i,j)$ \ (8)}
\\
& &&\left(\sum_{C \in \mathcal{C}_{\rm Large}} {\rm cn}^{(C)}y^{(C)}_{i,j}\right) \le N_{{\rm Large}} \ \tag*{$(\forall i,j)$ \ (9)} 
\\
& && \left(\sum_{C \in \mathcal{C} \setminus \mathcal{C}_{\rm Large}} {\rm cn}^{(C)}y^{(C)}_{i,j}\right) \le 
a^{*} \left(\sum_{C \in \mathcal{C}_{\rm Large}} {\rm cn}^{(C)}y^{(C)}_{i,j}\right) + b^{*} \ \tag*{$(\forall i,j)$ \ (10)}
\end{align}
\end{problem}

\section{Determining Delivery Route\label{sec:delivery_route}}

In this section, we determine the delivery route based on the following aspects with the input data: the division of customers, and every customer in the division is classified as a high-risk or moderate-risk customer obtained by the Problem~\ref{prob:multiBB}.
\begin{itemize}
    \item The customers are to be visited for delivery are determined among moderate-risk customers. We deliver to as many moderate-risk customers as possible with ensuring visit to high-risk customers.
    \item We determine the visiting order to the customers. Because MapBox\footnote{\url{https://www.mapbox.com}} is utilized to acquire the optimal route between two customers, the focus is only on the order of the visits. The working duration should be as short as possible.
\end{itemize}

Let $\bar{y}$ be the optimal solution of Problem~\ref{prob:multiBB}.
Because determining the delivery route is considered per day, the suffix $i$ is omitted. 
The sets of high-risk customers belonging to trip $j$ and that of moderate-risk customers belonging to trip $j$ are denoted as: $\mathcal{\hat{C}}_{\textrm{high}}(j) := \left\{C \in \mathcal{C}_{\textrm{high}}(i) \mid \bar{y}_{i,j}^{{(C)}} = 1 \right\}$ and $\mathcal{\hat{C}}_{\textrm{mdr}}(j) := \left\{C \in \mathcal{C}_{\textrm{mdr}}(i) \mid \bar{y}_{i,j}^{{(C)}} = 1 \right\}$, respectively.
Note that the indices of $j$ in $\mathcal{\hat{C}}_{\textrm{high}}{(j)}$ and $\mathcal{\hat{C}}_{\textrm{mdr}}{(j)}$ do not indicate the order of visiting per trip.

The delivery route is obtained by solving two optimization problems sequentially.
With the first problem, we determine a delivery route that maximizes the number of deliveries to moderate-risk customers, ensuring deliveries to high-risk customers.
When the two routes have the same number of deliveries, the routes can differ in terms of the working time.
Therefore, a second optimization problem is prepared to minimize the working time,
ensuring the replacement to the customers obtained from the first optimization problem.
When the solution is not found within the time limit, the heuristic algorithms shown in Section~\ref{sec:post_processing} are applied to determine the delivery route for applying the proposed system to the real world.

\subsection{Maximizing Number of Deliveries}
First, we obtain how many customers we can visit for replacement.
Note that high-risk customers must be visited for the replacement, and moderate-risk customers need not, but encouraged to be visited to suppress the concentration of the replacements.

\subsubsection{Graph Construction for Obtaining Multi-trip Delivery Order\label{sec:dr_graph}}

A graph is constructed for formulating a mixed-integer optimization problem.
The path on the graph represents the order of visiting.
When we denote the number of trips for the date as $n$, the sets of vertices and edges are prepared as follows:
\begin{enumerate}
    \item $V := SN \cup  \bigcup_{j=1}^{n}\left(\mathcal{\hat{C}}_{\textrm{high}}(j) \cup \mathcal{\hat{C}}_{\textrm{mdr}}(j) \right) $: Vertices
    \begin{itemize}
        \item $SN$: Set of supernodes representing the delivery center \\
    \end{itemize}
    \item $E := \cup_{j=1}^{n} \left(E^{\rm inter}_j \cup E^{\rm intra}_j \right)$: Edges
    \begin{itemize}
        \item $E^{\rm inter}_{j} \subset \left\{ (C,C^{\prime}) \in \left(\mathcal{\hat{C}}_{\textrm{high}}(j) \cup \mathcal{\hat{C}}_{\textrm{mdr}}(j) \right) \times \left(\mathcal{\hat{C}}_{\textrm{high}}{(j)} \cup \mathcal{\hat{C}}_{\textrm{mdr}}{(j)} \right) \mid C \neq C^{\prime}\right\}$. An edge $(C,C^{\prime})$ represents the movement of the staff from customer $C$ to customer $C^{\prime}$, both of which were selected during trip $j$. To reduce the number of variables, we extract the five-nearest-neighbor customers and create edges among the nodes corresponding to every customer.
        \item $E^{\rm intra}_j := \left(\mathcal{\hat{C}}_{\textrm{high}}(j) \cup \mathcal{\hat{C}}_{\textrm{mdr}}(j)\right) \times SN \cup  SN \times \left(\mathcal{\hat{C}}_{\textrm{high}}{(j)} \cup \mathcal{\hat{C}}_{\textrm{mdr}}{(j)} \right) $. An edge $(C,sn)$ represents the staff moving from customer $C$ to delivery center $sn$, and vice versa.
    \end{itemize}
    The trip index $j$ does not indicate the order of visiting, and the order of visiting per trip is determined after solving the optimization Problem~\ref{prob:maxDeli}.
\end{enumerate}

\subsubsection{Formulation}
We show the proposed formulation to maximize the number of deliveries as follows.
\begin{enumerate}
\item Variables
    \begin{itemize}
        \item $z^{(C, C^{\prime})}=
        \left\{
            \begin{array}{ll}
                1 & \text{Staff visit customer} \ C^{\prime} \ \text{next to} \ C \ \text{for replacement} \\
                0 & \text{otherwise}
            \end{array}
        \right.$
        \item $u^{(C)} \in \mathbb{Z}$: Order in which customer $C$ is visited during a particular day
        \item $t^{(C)} \in \mathbb{R}$: Arrival time to customer $C$ 
    \end{itemize}
\item Constants
    \begin{itemize}
        \item $d: V \times V \to \mathbb{R}$: Duration of movement between two points
        \item $\left[T_{lb}^{\rm work}, T_{ub}^{\rm work}\right] \subset [0, 24\times60]$ : Time window representing staff availability (minute)
        \item $T_{\rm max} $ : Sufficiently large value (at most $24 \times 60$)
        \item $\left[T_{lb}^{(C)}, T_{ub}^{(C)}\right] \subset \left[0,24\times60\right]$: Time window representing availability of customer $C$ 
        \item $rep: V \to \mathbb{R}$: Duration for replacing gas cylinder for customer, and duration for break time for supernode
        \item $n$: Number of trips
        \item $\mathcal{\hat{C}}_{\textrm{high}}(j)$: High-risk customers in trip $j$ selected by Problem~\ref{prob:multiBB}
        \item $\mathcal{\hat{C}}_{\textrm{mdr}}(j)$: Moderate-risk customers in trip $j$ selected by Problem~\ref{prob:multiBB}
        \item $f_j: (\mathcal{\hat{C}}_{\textrm{high}}(j) \cup \mathcal{\hat{C}}_{\textrm{mdr}}(j)) \rightarrow 2^{(\mathcal{\hat{C}}_{\textrm{high}}(j) \cup \mathcal{\hat{C}}_{\textrm{mdr}}(j))}$: Five-nearest customers from a customer in trip $j$
    \end{itemize}
\end{enumerate}

The formulation is shown in Problem~\ref{prob:maxDeli}.
The objective is to maximize the customers to visit for cylinder's replacement.
Constraints (11) and (12) represent the flow conservation for customers. 
Moreover, constraint (12) ensures visit to high-risk customers for replacement.
Constraints (13) and (14) ensure that every supernode satisfies the flow conservation and must be visited.
Constraint (15) is the subtour elimination constraint, and it restricts the relationship between variables $z$ and $u$.
Constraints (16) and (17) satisfy the time demands of the customers and staff, respectively.
Constraint (18) ensures the relationship between $z$ and $t$.
Constraint (19) ensures that the working time is included in the staff availability.

\begin{problem}
\label{prob:maxDeli}
\rm
Maximizing the visit to the moderate-risk customers for replacement
\begin{align}
& \underset{z,u,t}{\rm maximize} && \sum_{(C,C^{'})\in E} z^{\left(C,C^{'}\right)}  \notag \\
& \rm{subject \ to} && \sum_{C_{0} \in f_j(C)} z^{(C_{0}, C)}=\sum_{C_{1} \in f_j(C)} z^{(C, C_{1})} \leq 1 \  && & \tag*{$(\forall j, \forall C \in \mathcal{C}_{\textrm{mdr}}{(j)})$ \ (11)}
\\
& && \sum_{C_{0} \in f_j(C)} z^{(C_{0}, C)}=\sum_{C_{1} \in f_j(C)} z^{(C, C_{1})} = 1 \  && & \tag*{$(\forall j, \forall C \in \mathcal{C}_{\textrm{high}}{(j)})$ \ (12)}
\\
& && \sum_{C \in V \setminus SN} z^{(sn_1, C)} = \cdots = \sum_{C \in V \setminus SN} z^{(sn_{n-1}, C)} = 1 \   && & \tag*{$(sn_1 ,\ldots, sn_{n-1} \in SN)$ \ (13)}
\\
& && \sum_{C \in V \setminus SN} z^{(C, sn_2)} = \cdots = \sum_{C \in V \setminus SN} z^{(C, sn_{n})} = 1 \  && & \tag*{$(sn_2 ,\ldots, sn_{n} \in SN)$ \ (14)}
\\
& && u^{(C)} - u^{(C^{'})} + (|V|-1)z^{(C,C^{'})} \le |V| - 2 \  && & \tag*{$\left(\forall\left(C, C^{\prime}\right) \in E\right)$ \ (15)}
\\
& && T_{lb}^{(C)} \sum_{C^{\prime} \in V} z^{(C^{\prime}, C)} \leq t^{(C)} \leq \left(T_{ub}^{(C)}-rep({C})\right) \sum_{C^{\prime} \in V} z^{(C^{\prime}, C)} \quad && & \tag*{$(\forall C \in V \setminus SN)$ \ (16)}
\\
& &&T_{lb}^{\rm work} \le t^{(C)} \le T_{ub}^{\rm work} && & \tag*{$(\forall C)$ \ (17)}
\\
& && t^{(C^{\prime})}-t^{(C)} \geq\left(rep({C})+d({C, C^{\prime}})\right) z^{(C, C^{\prime})}-T_{\max }\left(1-z^{(C, C^{\prime})}\right) && & \tag*{$\left(\forall\left(C, C^{\prime}\right) \in E\right)$ \ (18)}
\\
& && t^{(C)} - t^{(C^{'})} \le T_{\rm max} - rep({C^{'}}) && & \tag*{$\left(\forall\left(C, C^{\prime}\right) \in E\right)$  \ (19)}
\end{align}
\end{problem}

\subsection{Minimizing Working Duration}
After solving problem~\ref{prob:maxDeli}, the second problem that obtains the delivery route with the shortest working duration is addressed.
Letting the optimal solution to Problem~\ref{prob:maxDeli} be $\bar{z}$, the customers to be replaced in trip $j$ are denoted as $\bar{\mathcal{C}}(j) := \{ C\in \mathcal{\hat{C}}_{\textrm{high}}(j) \cup \mathcal{\hat{C}}_{\textrm{mdr}}(j) |  \sum_{C^{\prime}  \in f(C)} \bar{z}^{(C, C^{\prime})} = 1 \}$.
We obtain the replacement route under the constraint of visiting all of these customers.

\subsubsection{Graph Construction for Obtaining Multitrip Delivery Order}

A graph is constructed for formulating a mixed-integer optimization problem.
The path on the graph also represents the order of visiting.
Because the basic concepts are the same as in Section~\ref{sec:dr_graph}, the detailed explanation is omitted in this section.
The sets of vertices and edges are prepared as follows:
\begin{enumerate}
    \item $\bar{V} :=SN\cup  \bigcup_{j=1}^{n}\bar{\mathcal{C}}{(j)} $: Vertices
    \item $\bar{E} := \cup_{j=1}^{n} \left(\bar{E}^{\rm inter}_i \cup \bar{E}^{\rm intra}_j \right)$: Edges
    \begin{itemize}
        \item $\bar{E}^{\rm inter}_{j} \subset \{ (C,C^{\prime}) \in \mathcal{\bar{C}}{(j)}\times \mathcal{\bar{C}}{(j)} \mid C \neq C^{\prime}\}$
        \item $\bar{E}^{\rm intra}_j := \mathcal{\bar{C}}{(j)}\times SN \cup  SN \times \mathcal{\bar{C}}{(j)} $
    \end{itemize}
\end{enumerate}

\subsubsection{Formulation}
We show the proposed formulation to minimize the working duration as follows.
\begin{enumerate}
\item Variables
    \begin{itemize}
        \item $\xi^{(C, C^{\prime})}=
        \left\{
            \begin{array}{ll}
                1 & \text{Staff visit customer} \ C^{\prime} \ \text{next to} \ C   \ \text{for replacement}  \\
                0 & \text{otherwise}
            \end{array}
        \right.$
        \item $\nu^{(C)} \in \mathbb{Z}$: Order in which customer $C$ is visited during a particular day
        \item $\tau^{(C)} \in \mathbb{R}$: Arrival time to customer $C$ 
    \end{itemize}
\item Constants
    \begin{itemize}
        \item $\bar{f}_j: \bar{C}(j) \rightarrow 2^{\bar{C}(j)}$: Five-nearest customers from a customer in trip $j$
    \end{itemize}
    Other notations are the same as those in Problem~\ref{prob:maxDeli}.
\end{enumerate}
The formulation is shown in Problem~\ref{prob:minWork}.
The objective function represents the total working duration, which should be minimized.
Constraint (20) represents the flow conservation for customers. 
Constraints (21) and (22) ensure that every supernode satisfies the flow conservation and must be visited.
Constraint (23) is the subtour elimination constraint, and it restricts the relationship between variables $\xi$ and $\nu$.
Constraints (24) and (25) satisfy the time demands of the customers and staff, respectively.
Furthermore, constraint (26) ensures the relationship between $\xi$ and $\tau$.
Constraint (27) ensures that the working time is included in the staff availability.

\begin{problem}
\label{prob:minWork}
\rm
Minimizing the total working duration
\begin{align}
& \underset{\xi,\nu,\tau}{\rm minimize} && \underset{C}{\max}(\tau^{(C)} + rep({C})) \notag \\
& \rm{subject \ to} && \sum_{C_{0} \in \bar{f_j}(C)} \xi^{(C_{0}, C)}=\sum_{C_{1} \in \bar{f_j}(C)} \xi^{(C, C_{1})} = 1 \  && & \tag*{$(\forall j,  \forall C \in \bar{C}(j))$ \ (20)}
\\
& && \sum_{C \in V \setminus SN} \xi^{(sn_1, C)} = \cdots = \sum_{C \in V \setminus SN} \xi^{(sn_{n-1}, C)} = 1 \   && & \tag*{$(sn_1 ,\ldots, sn_{n-1} \in SN)$ \ (21)}
\\
& && \sum_{C \in V \setminus SN} \xi^{(C, sn_2)} = \cdots = \sum_{C \in V \setminus SN} \xi^{(C, sn_{n})} = 1 \  && & \tag*{$(sn_2 ,\ldots, sn_{n} \in SN)$ \ (22)}
\\
& && \nu^{(C)} - \nu^{(C^{'})
} + (|\bar{V}|-1)\xi^{(C,C^{'})} \le |\bar{V}| - 2 \  && & \tag*{$\left(\forall\left(C, C^{\prime}\right) \in \bar{E}\right)$ \ (23)}
\\
& && T_{lb}^{(C)} \sum_{C^{\prime} \in V} \xi^{(C^{\prime}, C)} \leq t^{(C)} \leq \left(T_{ub}^{(C)}-rep({C})\right) \sum_{C^{\prime} \in V} \xi^{C^{\prime}, C} \quad && & \tag*{$(\forall C \in  \bigcup_{j=1}^{n}\bar{C}(j))$ \ (24)}
\\
& &&T_{lb}^{\rm work} \le \tau^{(C)} \le T_{ub}^{\rm work} && & \tag*{$(\forall C)$ \ (25)}
\\
& && \tau^{(C^{\prime})}-\tau^{(C)} \geq\left(rep({C})+d({C, C^{\prime}})\right) \xi^{(C, C^{\prime})}-T_{\max }\left(1-\xi^{(C, C^{\prime})}\right) && & \tag*{$\left(\forall\left(C, C^{\prime}\right) \in \bar{E}\right)$ \ (26)}
\\
& && \tau^{(C)} - \tau^{(C^{'})} \le T_{\rm max} - rep({C^{'}}) && & \tag*{$\left(\forall\left(C, C^{\prime}\right) \in \bar{E}\right)$  \ (27)}
\end{align}
\end{problem}

\subsection{Postprocessing\label{sec:post_processing}}
When a feasible solution cannot be obtained within the time limit, we stop solving the optimization problem using the solver.
Then, simple heuristic algorithms are used to obtain the replacement plan for any situation, which is essential for applying the proposed model to the real world.
The concepts of the heuristic algorithm are as follows.
\begin{itemize}
    \item The 2-opt algorithm acquires the shortest path for visiting all the given high-risk customers.
    \item The greedy algorithm extracts as many customers as possible for visiting based on the value of the risk function while satisfying staff availability.
\end{itemize}

\section{Field Test \label{sec:field_test}}

A field test was conducted in Chiba prefecture in Japan with more than 1,000 customers for approximately two weeks.
In the field test, the gas provider operated the replacement based on the output of our system.
Evaluation metrics were calculated after the field test.
The metrics for the field test were compared with those for the replacement record for the same season in the previous year.

\subsection{Experimental Settings \label{sec:FTsetting}} 

The hyperparameter settings in the field test are summarized in Table~\ref{tab:FTsetting}.
Gurobi Optimizer v9.1.0 was utilized to solve the mixed-integer optimization problems, and the time limit for computing was set to 30 min.
Because the system was operated in a cloud computing environment in the field test, it was not possible to determine which computational environment was utilized per problem. 
The candidates are shown in the Appendix~\ref{sec:ft_computation}.
\begin{table}[tbp]
    \centering
    \begin{tabular}{c|c|c|c|c}
        $\alpha_{\textrm{high}}, q_{\textrm{high}}$ & $\alpha_{\textrm{mdr}}, q_{\textrm{mdr}}$ & $D_{lbd}$ & $D_{ph}$  & Working time \\ \hline
        $0 \%, 0.9$ & $5 \%, 0.1$ & $3$  & $3$ & 7:00 -- 19:00 
    \end{tabular}
    \caption{Hyperparameter settings in the field test. 
    Because the experiments described in Appendix~\ref{sec:appendix_dldr} were conducted after the field test, hyperparameter tuning was not performed in the field test. Note that the staff worked only on weekdays.}
    \label{tab:FTsetting}
\end{table}

\begin{wrapfigure}[17]{r}[1mm]{0.4\linewidth}
\begin{center}
\includegraphics[width=0.82\linewidth]{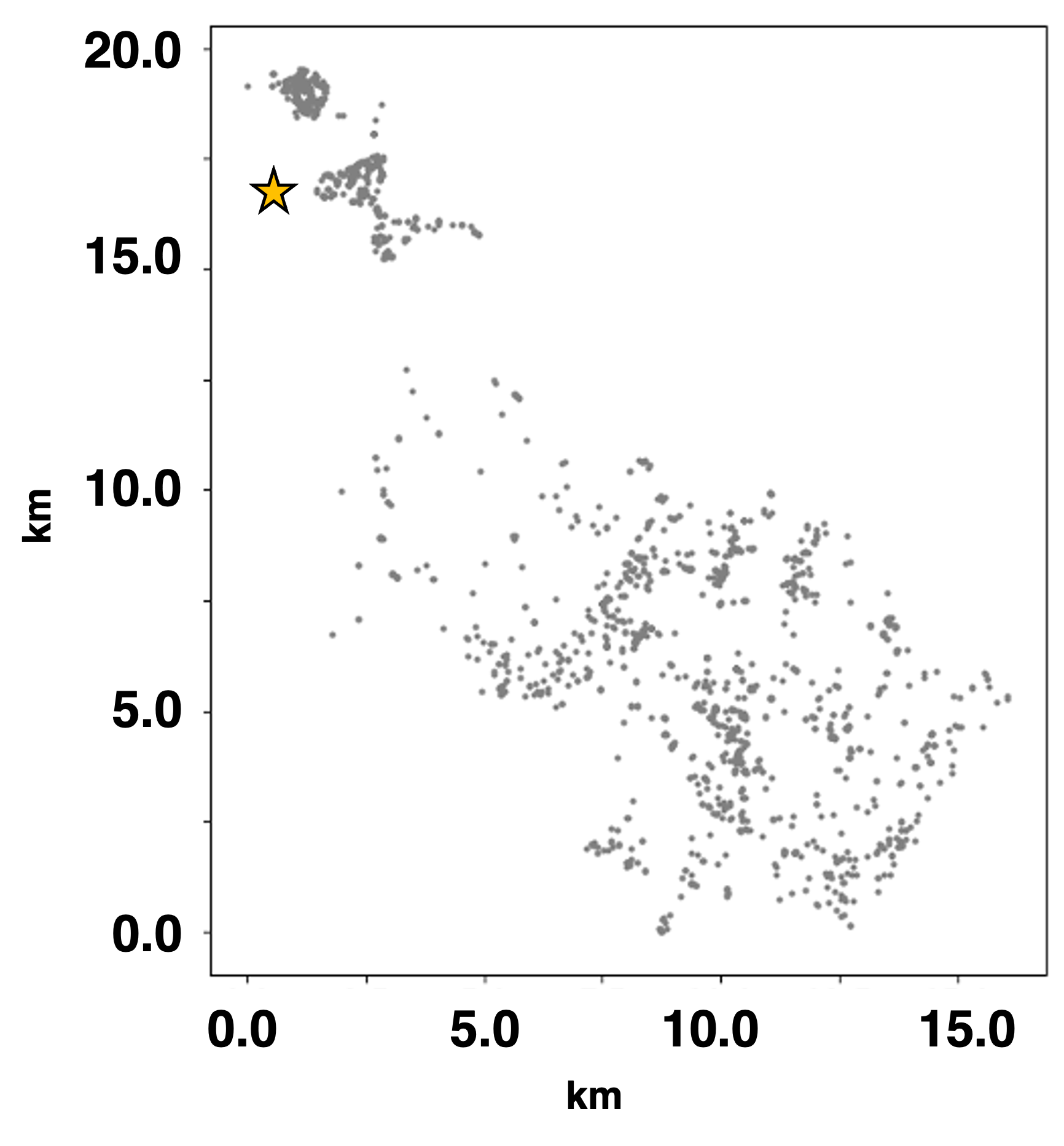}
\caption{Scatter plot of the customer locations and the delivery center (depot location)}
\label{fig:location2}
\end{center}
\end{wrapfigure}
The total number of customers was 1,366, and the number of customers with smart meters was 861.
The locations of the customers and the delivery center are shown in Figure~\ref{fig:location2}.
In the figure, the dots and the star represent the customers and the delivery center, respectively.
For the target date, the replacement plan for the next working date was created.
The driver works Monday through Friday.
The driver loads the cylinders on the truck after the delivery of the previous day, but loads the cylinders to be delivered on Monday after delivery on Friday. 
Our system was allowed to run for five hours, and the calculation was performed on the business day immediately preceding the delivery date.

In the gas usage forecast for estimating replacement dates, the support vector regression was used for the smart meter based on the pre-experiment. 
The proposed method (Section~\ref{sec:demand_forecast_sub}) was applied for most of conventional meter's gas usage estimation based on the pre-experiment shown in Appendix~\ref{sec:exp_dp}.

Based on the knowledge of the gas provider, the replacement plan operated about one year before the field test was picked up to evaluate the proposed system.
Because the gas usage heavily depends on the temperature, the comparison periods were extracted based on the average temperature.
The temperature data were obtained from the Japan Meteorological Agency website\footnote{\url{http://www.data.jma.go.jp/obd/stats/etrn/index.php}}.

\subsection{Evaluation Metrics \label{sec:FT_metric}}
The quality of the replacement plan was investigated using the following evaluation metrics.
\begin{itemize}
    \item Average traveling time per day (travel-time/day)
    \item Average traveling time per customer (travel-time/customer)
    \item Total number of nonreplacement visits (nonreplacement)
\end{itemize}
These metrics are different from those described in Section~\ref{sec:exp_dldr}.
``Travel time per day'' and ``Travel time per customer'' do not include the time required to replace the cylinders at the customers' locations and nonreplacement customers who are visited, that is, the driving time of the truck.
The abbreviations in the parentheses indicate the evaluation metrics listed in the Table~\ref{tab:FT_result_edited}.
The average number of customers whose gas has run out cannot be compared because of missing data.

\subsection{Results and Discussion}
The evaluation metrics are shown in Table~\ref{tab:FT_result_edited}, and the observations made are as follows.

\renewcommand{\arraystretch}{1.2}
\begin{table}[tbp]
    \centering
    \begin{tabular}{c|c|c}
        Name                    & Gas provider        & Proposed system      \\ \hline
        Period                  & 13/3/2020-27/3/2020 & 1/3/2021-15/3/2021  \\
        Average temperature     & 9.7 $^\circ$C        & 9.8 $^\circ$C        \\ \hline \hline
        Travel-time/day(↓)      & 2 h 52 m 32 s            & \textbf{2 h 06 m 44 s}   \\
        Travel-time/customer(↓) & 3 m 46 s               & \textbf{2 m 34 s}      \\
        Nonreplacement(↓)      & 78                  & $\textbf{0}^{*1}$ \\
    \end{tabular}
    \caption{Experiment setting and evaluation metrics of the field test.
    Owing to the missing data of the amount of gas remaining, it was not possible to evaluate gas shortage.
    Note that the gas provider occasionally changed the proposed system outputs, utilizing $83.9\%$ of the output of the algorithm. 
    }
    \label{tab:FT_result_edited}
\end{table}

\begin{itemize}
    \item The number of nonreplacement visits was determined to be zero.
    The staff was penalized by the gas provider when the average remaining gas rate at the replacement exceeds 8\% per month. 
    During the field test, to determine the number of cylinders that do not require nonreplacement based on the remaining gas rate, the gas provider was asked to replace the gas cylinder even when a large amount of gas remained.
    After the field test was finished, the average remaining gas rate was below 8\%.
    Therefore, it was determined that there were no nonreplacement visits, although the number of visits to customers in the proposed system was greater than that of the gas provider (see *1 in Table~\ref{tab:FT_result_edited}).
    \item The proposed system reduced the duration of travel-time and the number of nonreplacement events.
    It shows that the proposed system effectively determines the customer list and delivery route.
\end{itemize}


\section{Verification Experiments \label{sec:exp_dldr}}
After the field test, further experiments were conducted to investigate the influence of the replacement plan for each part of the proposed method.
The entire process of solving the CRP was divided into two parts: 1) estimating the date for replacement and 2) acquiring the customer list and determining the delivery route.
Because each of them is performed by the gas provider or the proposed system, four types of experiments were conducted.
\subsection{Experimental Settings \label{sec:dldr_experiment_setting}}
The basic hyperparmeter settings were as same as in Section~\ref{sec:FTsetting}.
These settings are shown in the upper part of Table~\ref{tab:exp_ISG_result_edited}.
The target dates were from March 16–24, 2021.

The gas usage forecast model of the gas provider used in Exp A and Exp C had the following features.
\begin{itemize}
    \item The daily gas usage was assumed to be constant.
    \item Five different periods were prepared, and the average value was calculated per period. Then, the maximum value among them was obtained.
    \item The forecast value was calculated by multiplying the above value and the hyperparameter representing the usage tendency of the next period.
\end{itemize}
Moreover, the workers manually created the replacement plan based on their experience with Exp A and B.
Based on the result of pre-experiment shown in Appendix~\ref{sec:appendix_dldr}, the hyperparameters were set, $\alpha_{\textrm{high}}$ as $0\%$, $q_{\textrm{high}}$ as $0.3$, $\alpha_{\textrm{mdr}}$ as $7\%$, and $q_{\textrm{mdr}}$ as $0.3$, respectively.

Gurobi Optimizer v9.1.1 was used to solve the mixed-integer optimization problems, and the time limit was set to 30 min.
In the computing environment, the CPU was an Intel(R) Xeon(R) Gold 6240R CPU with a 2.40GHz CPU frequency, and the memory was 768GB.

\subsection{Evaluation Metrics \label{sec:metric}}
The following evaluation metrics were calculated to investigate the quality of the replacement plan.
\begin{itemize}
    \item Average remaining gas rate at replacement (rate-average)
    \item Median of remaining gas rate at replacement (rate-median)
    \item Total success replacement rate, excluding failures (success)
    \item Rate of failure in replacement owing to a cylinder being out of gas (fail-out)
    \item Rate of failure in replacement owing to an overabundance of gas where the threshold was set as 15\%. (Fail-over) Note that the fail-over is different from the nonreplacement visit because nonreplacement visit is determined by staff’s sense. Therefore, the fail-over metric is calculated for quantitative evaluation.
    \item Number of failures in replacement owing to the time window of the customer (fail-time)
    \item Average delivery time per customer (time/customer)
    \item Average travel distance per customer (distance/customer)
    \item Average number of customers whose gas has run out (run-out)
\end{itemize}
The delivery time included the duration of replacing the cylinders in addition to the driving duration.
The abbreviations in parentheses are used to summarize the results listed in Table~\ref{tab:exp_ISG_result_edited}.

\subsection{Results and Discussion}
We summarize the results of the evaluation metrics in Table~\ref{tab:exp_ISG_result_edited}.
In summary, when the proposed system was utilized for the entire process of making replacement plans (Exp D), the highest replacement plan quality was achieved.
This was because Exp D achieved the best run-out result and second-best fail-over result.
Moreover, it succeeded in creating efficient replacement plans that can be expressed through the time/customer or distance/customer.
Because approximately one-third of customers did not have a smart meter in the experiment, these results indicate that the proposed system can make the replacement plan more effective even if not all customers have a smart meter.
The following aspects are considered based on the results.

\begin{figure}
    \centering
    \includegraphics[width=0.8\linewidth]{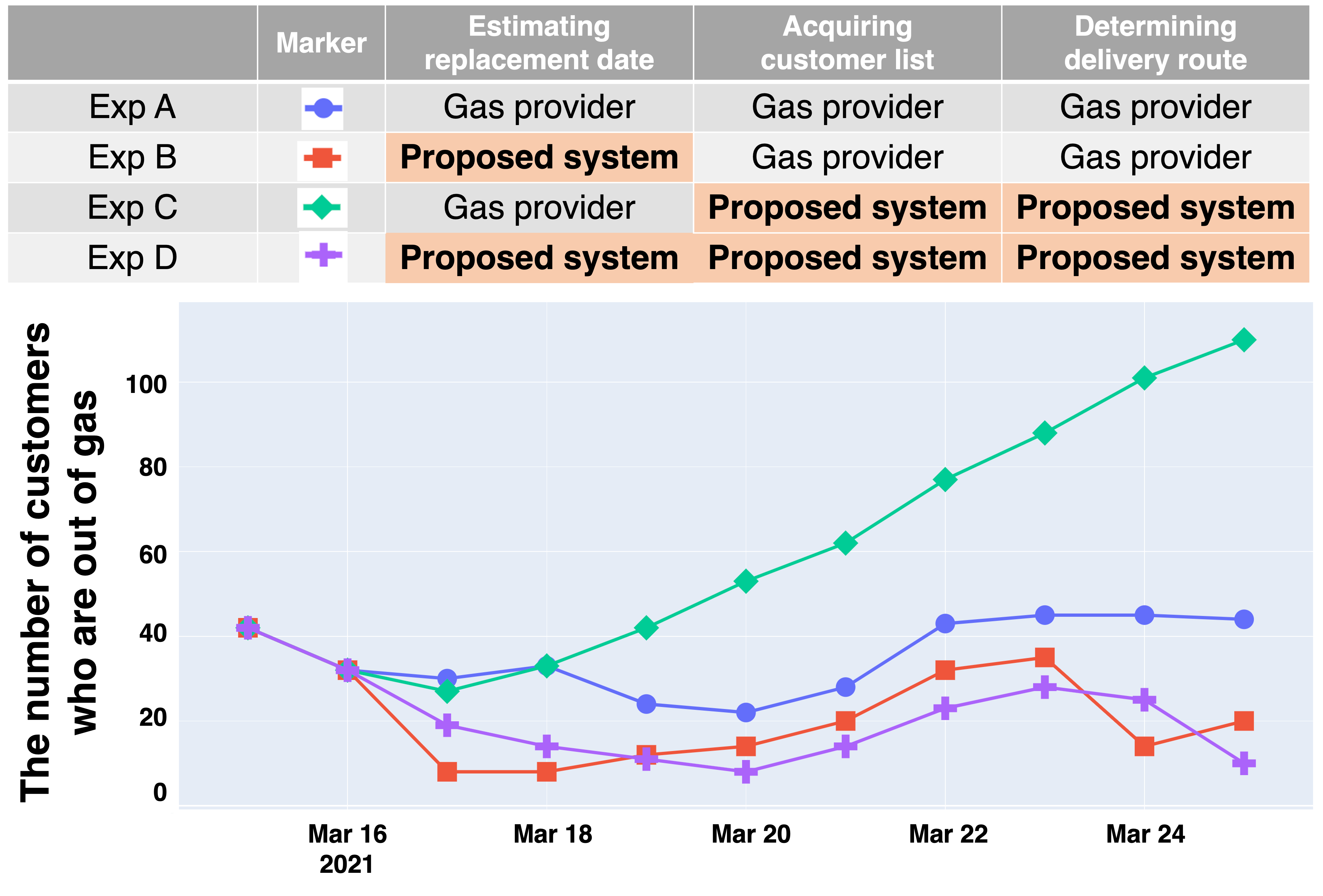}
    \caption{Number of customers whose gas has run out during the target period. Note that the lines overlap if the numbers are the same for the first two consecutive days.}
    \label{fig:outLCs}
\end{figure}

\begin{enumerate}
    \item (\textbf{Overall}) In Exp D, the proposed system was completely applied to make replacement plans.
    \begin{itemize}
        \item Exp D was more successful than Exp A in terms of all evaluation metrics.
        Exp A recorded the least number of fail-out visits; however, this was mainly because they left out the existence of customers with a gas shortage (see Figure~\ref{fig:outLCs}).
	    By contrast, the figure shows that
	    in Exp D, more high-risk customers were visited. 
        These tendencies can be observed by comparing the run-out events between Exp A and Exp D.
        \item Exp D achieved a smaller number of customers with run-out events and competitive results in fail-over events compared to Exp B.
        It indicates that the proposed system can realize a better replacement plan than the gas provider.
        Moreover, in Exp D, considerably better performance for time/customer and distance/trip was achieved than those in Exp B.
        We prepared for a future increase in high-risk customers by visiting  moderate-risk customers, which can be interpreted as the slope occurring near March 22, as shown in Figure~\ref{fig:outLCs}.
        The figure shows that Exp D achieved the smallest increase in the number of customers experiencing a gas shortage.
        In other words, a larger number of failure events indicates that the visit for replacement was assured to the customer after gas ran out.
        Because the rate-median of Exp D was lower than that of Exp A and Exp C, Exp D tended to achieve visit to customers for replacement who were about to run out of gas.
        \item Through the proposed system, more efficient replacement planning was realized than in Exp C for every evaluation metric. 
        The number of run-out events could be dramatically reduced in Exp D based on a better gas usage forecast.
        \item In Exp D, more fail-out visits were observed. 
	By contrast, Exp D succeeded in reducing the number of run-out events compared with Exp A.
        Exp D focused more on delivery to high-risk customers, which caused a decrease in the accumulation of run-out events.
        The smallest rate-median was reported in Exp D, indicating that the focus was on replacement high-risk customers' cylinders during Exp D.
    \end{itemize}
    \item (\textbf{Estimating replacement date}) The proposed system was compared with a conventional method used for estimating replacement date. 
    In both Exp A and B, the replacement plan was made by the gas provider. 
    In Exp A, the gas provider executed the gas usage forecast, whereas the proposed system in Exp. B conducted.
    \begin{itemize}
        \item Rate-median in Exp B was lower than in Exp A. 
        With the proposed system for gas usage forecasting, the gas provider could deliver to customers with a low remaining gas rate. 
        The results show the superiority of the proposed system in terms of estimating the replacement date.
    \end{itemize}
    \item  (\textbf{Acquiring customer list and determining delivery route}) The proposed system and the conventional system for acquiring a customer list and determining a delivery route were compared. Based on the gas usage forecast result of the gas provider, the replacement plan was made by the gas provider and the proposed system during Exp A and Exp C, respectively.
    \begin{itemize}
        \item The distance/trip in Exp C was longer than that in Exp A.
        Because the proposed system focuses on visits to high-risk customers with the highest priority, the delivery route occasionally becomes longer than the replacement plan of the gas provider.
        Exp C recorded the most run-out events owing to the accuracy of the gas usage forecast. 
        \item In Exp C, as shown in Figure~\ref{fig:outLCs}, the number of customers with run-out events increased. 
        Many customers were categorized as high-risk customers owing to the imprecise gas usage forecast.
	    This caused many high-risk customers and some moderate-risk customers, who were also considered high-risk customers.
        When there were too many trips, some trips were extracted to calculate the route for replacement.
        Based on the results, there were the categorization error that high-risk consumers actually had enough gas and should not be replaced. 
        Thus, the number of moderate-risk customers increased over time.
        It implies that the quality of the replacement
        plan can be improved by modifying the selection of many high-risk customers, which is planned as future research.
    \end{itemize}
\end{enumerate}

\renewcommand{\arraystretch}{1.0}
\begin{table}[htbp]
    \centering
    \begin{tabular}{c|c|c|c|c}
        Name            & Exp A           & Exp B           & Exp C          & 
        \begin{tabular}{c}
             Exp D  \\
             (Exp 3 in \\ 
             Appendix~\ref{sec:appendix_dldr})
        \end{tabular}
        \\ \hline
        \begin{tabular}{c}
            Estimating date \\ for replacement  
        \end{tabular} & Gas provider & Proposed        & Gas provider   & Proposed           \\ \hline
        \begin{tabular}{c}
        Acquiring \\ customer list       
        \end{tabular}
        & Gas provider & Gas provider & Proposed   & Proposed           \\ \hline
        \begin{tabular}{c}
        Determining \\ delivery order
        \end{tabular} & Gas provider & Gas provider    & Proposed       & Proposed           \\ \hline \hline
        Rate-average          & 10.3 \%      & 8.83e-05 \%     & 8.73 \%        & -1.36 \%           \\
        Rate-median           & 7.56 \%      & 1.82 \%         &  7.07 \%       & 0.09 \%            \\
        Success               & 36.3 \%      & 56.6 \%         & 31.3 \%        & 47.1 \%            \\
        Fail-out              & 30.5 \%      & 38.3 \%         & 35.4 \%        & 45.1 \%            \\
        \textbf{Fail-over(↓)} & 33.2 \%      & \textbf{4.7 \%} & 33.3 \%        & 7.8 \%             \\
        Fail-time (↓)          & 3            & 2               & \textbf{0}     & \textbf{0}         \\ 
        Time/customer(↓)      & 12 m 55 s    & 13 m 30 s       & 13 m 19 s      & \textbf{12 m 05 s} \\
        Distance/customer(↓)     & 1.7 km      & 2.7 km         & 2.7 km        & \textbf{1.5 km}   \\ 
        \textbf{Run-out(↓)}   & 34.4         & 21.7            & 49.7           & \textbf{21.6}
    \end{tabular}
    \caption{Settings and evaluation metrics of the experiment. 
    As the criterion for failure, the remaining gas rate at replacement must be greater than 15\%. 
    The results of Exp D were the same as those in Exp 3 listed in Table~\ref{tab:exp_result_edited} shown in Appendix~\ref{sec:appendix_dldr}.}
    \label{tab:exp_ISG_result_edited}
\end{table}

\section{Conclusion\label{sec:conclusion}}
In this paper, we consider the CRP that aims to create a plan to visit customers for replacing gas cylinders, including the delivery route.
One of the difficulties for CRP derives from the data acquisition of data on the remaining gas because the gas provider cannot determine the remaining gas without visiting customers and observing the meter.
To address this problem, smart meters can be installed for customers, enabling the gas provider to acquire gas usage data daily.
In this study, we construct a system to solve CRP.
Our system comprises three steps: estimating the date for replacement, acquiring the customer list, and determining the delivery route.
Our formulation for making the customer lists enables the gas provider to replace cylinders some days before the date about to be a gas shortage.
It can suppress the concentration of replacement on certain days.
Numerical experiments were conducted for real gas usage in Chiba prefecture in Japan with more than 1,000 customers.
A field test was performed by a gas provider incorporating the output of the proposed system into its operation.
Our system succeeded in reducing the gas shortage, nonreplacement visits when an ample supply of gas remained, and working duration.
Note that because approximately one-third of customers did not have a smart meter, the results show the potential of utilizing smart meters.
As more smart meters are installed to the customer, the gas shortage will decrease because of the acquisition of more-precise replacement date estimation.

Three research directions for the future work are being considered.
After the field test, we got feedback from the gas providers that the customer for replacement per day should be more aggregated.
Based on the feedback, how to acquire the customer list so that the customers for replacement are more close to each other will be investigated as the first topic.
When this issue is addressed, for example, gas providers can replace cylinders for multiple customers without changing parking positions, which reduces the staff workload more effectively.
The second direction for future work is how to determine customer priority for installing smart meters.
In this study, the cylinder replacement problem after installing the smart meter for some customers was considered.
In the future, our system will be utilized to make replacement plans for other areas where no smart meters are installed.
In that case, it is necessary to determine the customer's priority for installing smart meters to improve the replacement date's estimation performance.
The third direction is utilizing our framework for different industries.
Our system creates optimal delivery plans through suppressing the sudden increase in delivery targets, consisting of demand forecasting, acquiring customer list for delivery, and determining the delivery route.
The proposed system can be applied to solve related problems of other industries, such as grocery delivery.

\section*{Acknowledgement}


This research project was supported by the Japan Science and Technology Agency (JST), the Core Research of Evolutionary Science and Technology (CREST), the Center of Innovation Science and Technology based Radical Innovation and Entrepreneurship Program (COI Program), JSPS KAKENHI Grant Number JP16H01707 and JP21H04599, Japan.
It is also supported by SoftBank Corp. and ISG, Inc..

\bibliographystyle{unsrt}
\bibliography{main}

\clearpage
\vspace{15pt}
{\leftskip 8.5cm \parindent 0mm
Akihiro Yoshida\\
Graduate School of Mathematics, Kyushu University\\
744, Motooka, Nishiku, Fukuoka, 819-0395, Japan\\
E-mail: \texttt{akihiro.yoshida.916@kyudai.jp}
\par}
\appendix
\section*{Appendix}
\section{Computational Environment Used in Field Test \label{sec:ft_computation}}
In this section, the computation environment used in field test is described (see Section~\ref{sec:field_test}).
Different computing environments were prepared for estimating the replacement date and acquiring customer lists, and determining delivery routes.
One of the following computational environments with 8-GB memory was utilized for estimating the replacement date.
\begin{itemize}
    \item Intel(R) Xeon(R) Platinum 8370C with 2.1-GHz CPU frequency
    \item Intel(R) Xeon(R) Platinum 8272CL with 2.1-GHz CPU frequency
    \item Intel(R) Xeon(R) Platinum 8171M 2.1-GHz CPU frequency
    \item Intel(R) Xeon(R) E5-2673 v4 with 2.3-GHz CPU frequency
    \item Intel(R) Xeon(R) E5-2673 v3 with 2.4-GHz CPU frequency
\end{itemize}
Moreover, one of the following computational environments with 128-GB memory was utilized for acquiring customer list'' and determining delivery route.
\begin{itemize}
    \item Intel(R) Xeon(R) Platinum 8370C with 2.1-GHz CPU frequency
    \item Intel(R) Xeon(R) Platinum 8272CL with 2.1-GHz CPU frequency
    \item Intel(R) Xeon(R) 8171M with 2.1-GHz CPU frequency
    \item Intel(R) Xeon(R) E5-2673 v4 with 2.3-GHz CPU frequency
\end{itemize}
\section{Numerical Experiment to Estimate the Replacement Date \label{sec:exp_dp}}
In this section, we explain the experiment for estimating the replacement date.
Two experiments were conducted.
The first focused on evaluating the forecast performance for the gas usage of a conventional meter.
The second aimed to investigate the influence of hyperparameters for extracting the high-risk customers by considering the gas usage forecast's uncertainty.
We denote proposed model shown in Section~\ref{sec:demand_forecast_sub} as kNN.
\subsection{Evaluation of Gas Usage Forecast for Conventional Meters}
\label{sec:exp_forecast}
The method of the gas usage forecast presented in Section~\ref{sec:demand_forecast_sub} was evaluated using a conventional meter. 

\subsubsection{Evaluation Settings}
For the experiment conducted with conventional meters, smart meter data were used to prepare the correct data.
Data were generated by intentionally omitting the smart meters' data every 29 days, which is the maximum frequency of the missing data of meters for conventional meters.
\begin{itemize}
    \item Target dates are 42 days: $\{1/2/2021, \cdots, 14/3/2021\}$
    \item Target meters : 619 smart meters
    \item Number of samples $k$  : 10
\end{itemize}

\subsubsection{Evaluation Metrics}
The root-mean-square error (RMSE) of each meter was calculated for evaluation.
In addition, the {\it upward rate} was proposed as an evaluation metric. 
To avoid running out of gas, it is important that the forecast gas usage be less than the actual gas usage.
Therefore, the {\it upward rate} is defined as the percentage of ${\textit p
rediction} \geq {\textit observation}$, and it was determined that a higher {\it upward rate} was better.
\subsubsection{Baselines}
Four baseline models were prepared.
\begin{itemize}
    \item EachMean: It uses the mean of its own meter data over the past 90 days as the forecast value. 
    \item TQ: It uses the third quartile $Q_3$ of its own meter data over the last 90 days as the forecast value. 
    \item AllMean: It uses the mean of all smart meter data over the past 90 days as the forecast value. 
    \item MaxModel : It imitates the gas usage forecast model of the gas provider that is described in Section~\ref{sec:dldr_experiment_setting}.
\end{itemize}

\subsubsection{Results and Discussion}

\begin{wrapfigure}[16]{r}[3mm]{0.5\linewidth}
    \centering
    \includegraphics[width=7cm]{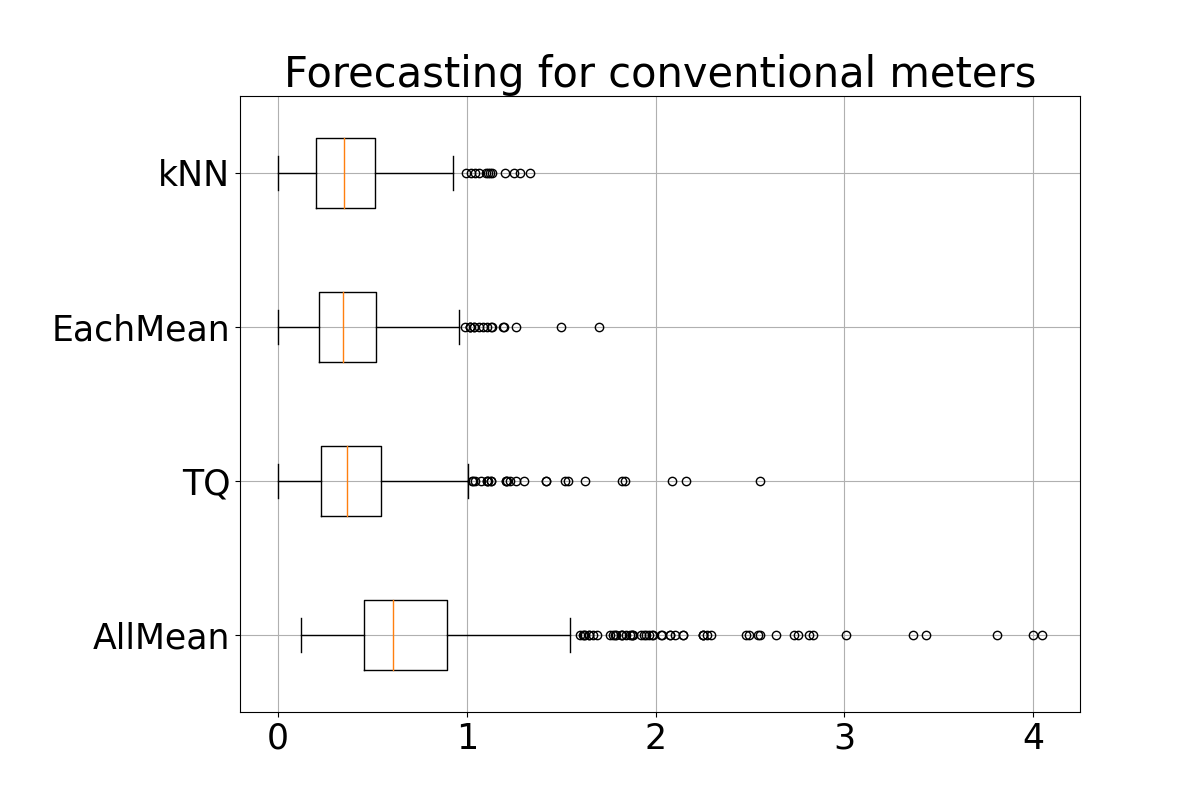}
    \caption{Box plot of RMSE values in the gas usage forecast. MaxModel is eliminated because its performance was much worser than the others.}
    \label{fig:conventional}
\end{wrapfigure}

The experimental results for conventional meters are presented in Figure~\ref{fig:conventional} and in Table~\ref{tab:conventional}.
\begin{enumerate}
    \item (MaxModel versus kNN/EachMean/TQ/AllMean) The average RMSE values among customers of the proposed models were smaller than those of the baseline Max model.
    \item (kNN versus EachMean/TQ/AllMean) The RMSE values of the kNN were smaller than those of the other models.
    This indicates that similar meters data enables us to obtain the more precise gas usage forecast compared to simple methods.
\end{enumerate}

The kNN uses smart meter data in addition to its own conventional meter value.
By contrast, Eachmean and TQ uses only its own conventional meter value, and All Mean uses only the mean of all smart meter data.
The numerical results show that the kNN model performs better than the EachMean, TQ, and All Mean models.
In other words, utilizing the smart meter's gas usage enhances the performance.
However, when the gas usage of the conventional meter is sufficiently larger than that of the other meters, the kNN model does not perform better. 
This is because the forecast value may be smaller than actual daily gas usage. 
Under this situation, the TQ model is the best choice because it has the highest upward rate.
Therefore, in our experiment, for the gas usage forecast, kNN was applied for most of conventional meters, and TQ was applied for the others with unique gas usage tendency such as a hotel or a restaurant.

\begin{table}[]
\centering
\begin{tabular}{crrrrr}
\multicolumn{1}{c|}{Method} & \multicolumn{1}{c|}{Mean}   & \multicolumn{1}{c|}{Median} &  \multicolumn{1}{c|}{IQR} & \multicolumn{1}{c|}{Maximum} & \multicolumn{1}{c}{Upward rate} \\ \cline{1-6}
\multicolumn{1}{c|}{MaxModel}   & \multicolumn{1}{r|}{1.3628} & \multicolumn{1}{r|}{0.4227} & \multicolumn{1}{r|}{0.3902} & \multicolumn{1}{r|}{2.165e+02} & \multicolumn{1}{r}{0.5010} \\ 
\multicolumn{1}{c|}{kNN (Proposed)}   & \multicolumn{1}{r|}{\textbf{0.3795}} & \multicolumn{1}{r|}{0.3478} & \multicolumn{1}{r|}{0.3108} & \multicolumn{1}{r|}{\textbf{1.3350}} & \multicolumn{1}{r}{0.5518} \\ 
\multicolumn{1}{c|}{EachMean}   & \multicolumn{1}{r|}{0.3895} & \multicolumn{1}{r|}{\textbf{0.3440}} & \multicolumn{1}{r|}{\textbf{0.2990}} & \multicolumn{1}{r|}{1.7005} & \multicolumn{1}{r}{0.5916} \\ 
\multicolumn{1}{c|}{TQ}   & \multicolumn{1}{r|}{0.4243} & \multicolumn{1}{r|}{0.3632} & \multicolumn{1}{r|}{0.3176} & \multicolumn{1}{r|}{2.5515} & \multicolumn{1}{r}{\textbf{0.6708}} \\ 
\multicolumn{1}{c|}{AllMean}   & \multicolumn{1}{r|}{0.7722} & \multicolumn{1}{r|}{0.6084} & \multicolumn{1}{r|}{0.4431} & \multicolumn{1}{r|}{4.0473} & \multicolumn{1}{r}{0.4489} \\ 
\end{tabular}
\caption{Evaluation metrics for each forecast model for conventional meters. 
Mean/Median/IQR/Maximum indicate the mean, median, interquartile range, and maximum of the RMSE values, respectively, in which a lower value is better. 
}
\label{tab:conventional}
\end{table}

\subsection{Parameter Sensitivity for Extracting High-Risk Customers}

In this section, the effect of the threshold to determine high-risk customers is examined based on the risk functions value shown in Section~\ref{sec:risk_func}.
\begin{figure}
    \centering
    \includegraphics[width=0.85\linewidth]{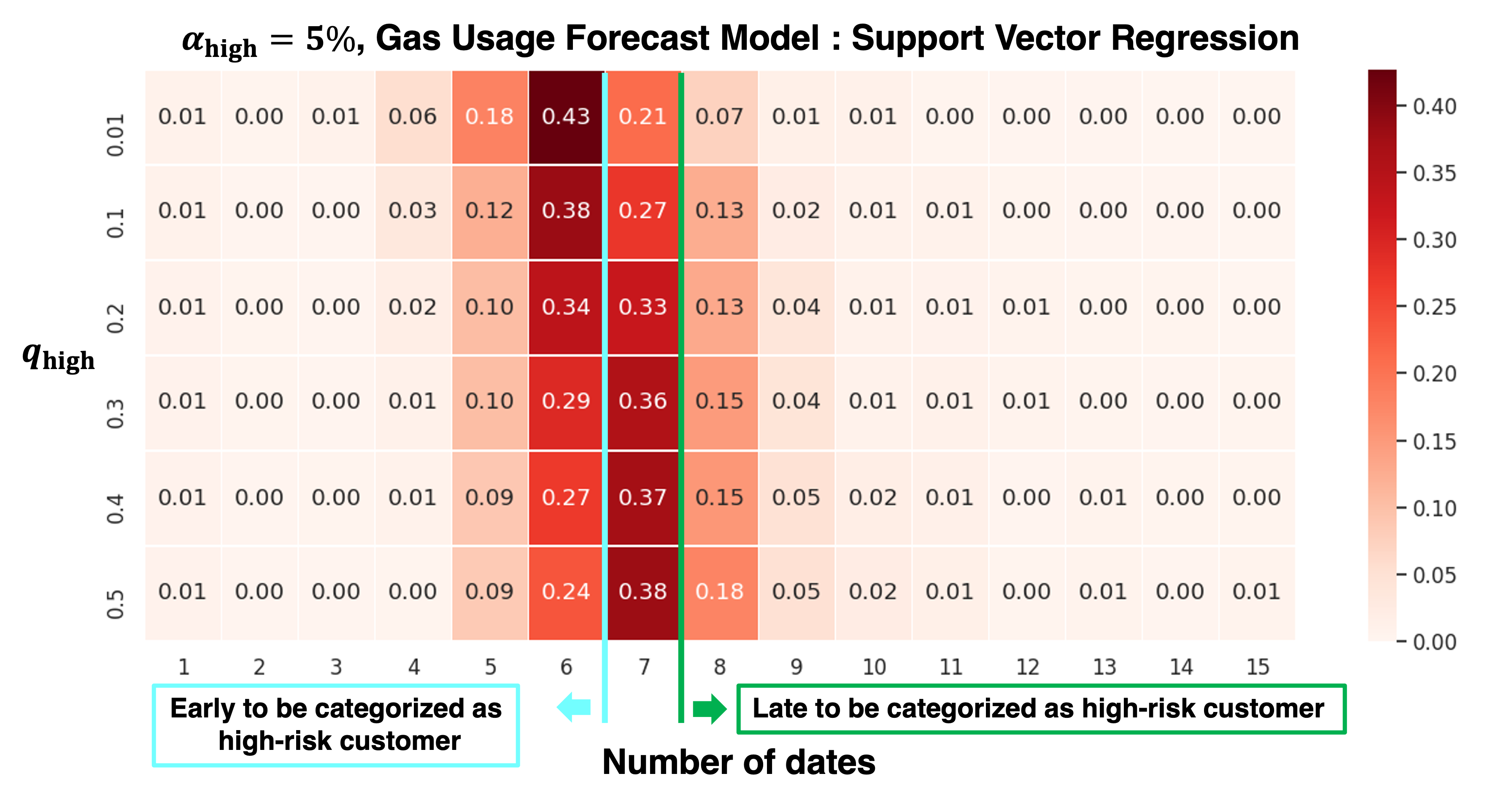}
    \caption{Prediction results for categorizing high-risk customers based on gas usage forecast per day using support vector regression when changing the threshold $q_{\textrm{high}}$}
    \label{fig:predhard_exp1_svr}
    \centering
    \includegraphics[width=0.85\linewidth]{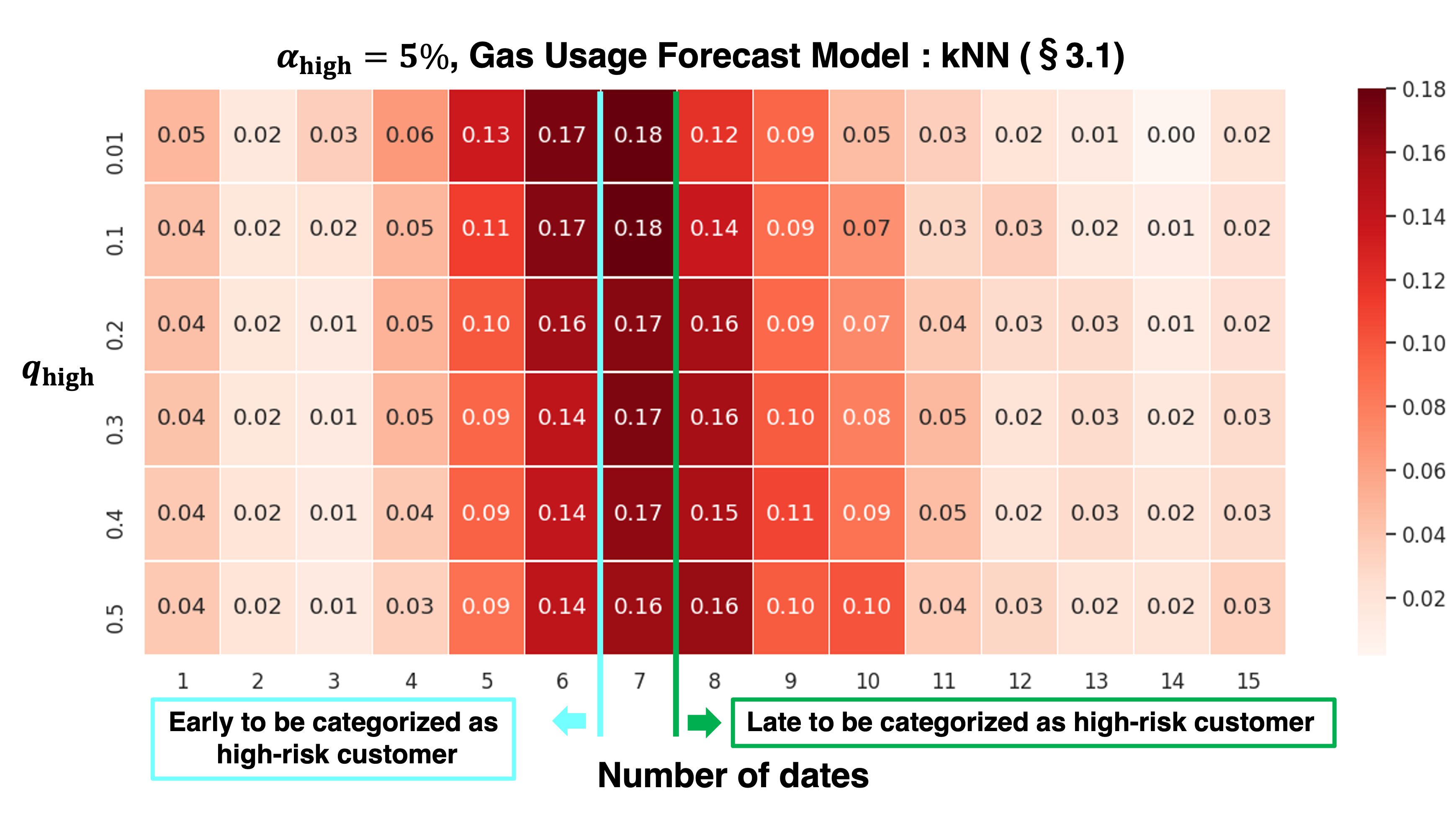}
    \caption{Prediction results for categorizing high-risk customers based on gas usage forecast per day using proposed kNN when changing the threshold $q_{\textrm{high}}$}
    \label{fig:predhard_exp3_kNN}
\end{figure}
\subsubsection{Evaluation Settings}
We extracted meters with no missing values from April 1, 2020 to March 23, 2021.
The dates from when the remaining gas rate decreased below the threshold to the remaining gas rate in the date units were determined. 
Then, starting from the date seven days preceding the gas usage forecast, the risk function was used to calculate when the customer becomes a high-risk customer.
During the experiment, the threshold of the gas rate was set to 5\%.
Under certain settings, the meter data were intentionally removed by utilizing real missing data.
The following two settings were prepared.
\begin{itemize}
    \item Exp 1: The complete data were utilized without intentionally introducing missing data. 
    Because the setting is assumed for customers with smart meters, we use the support vector regression.
    \item Exp 2: The data were intentionally removed every 29 days to reproduce the data acquisition of conventional meters. The kNN described in Section~\ref{sec:demand_forecast_sub} was used.
\end{itemize}

\subsubsection{Results and Discussion}

The results of the experiment on the change in threshold in Exp 1 and Exp 2 are shown in  Figures~\ref{fig:predhard_exp1_svr} and ~\ref{fig:predhard_exp3_kNN}, respectively.
Some customers were predicted to be high-risk with an error of one day with respect to the correct days, seven days.
To address this situation, the customer can be categorized as a high-risk customer earlier by decreasing the threshold $q_{\textrm{high}}$.
It was experimentally observed that the distribution of the days predicted to be a high-risk was shifted earlier by one day with a threshold $q_{\textrm{high}}$ of 0.01. 
Under this situation, the rate of customer prediction to be later than the actual date (particularly after seven days) decreases from 28\% to 10 \% when $q_{\textrm{high}}$ changes from 0.5 to 0.01, respectively.
Customers were categorized as high-risk customers earlier as the threshold decreased.
Although users can freely choose the threshold value based on these figures, a threshold of $0.3$ was chosen for the experiment in Appendix~\ref{sec:appendix_dldr} because it was the lowest value among the values where the most frequent value was seven.


\section{Hyperparameter Tuning in Making Customer List \label{sec:appendix_dldr}}
In this section, we investigate the influence of the parameter settings for making the customer list and determining the delivery route.
\subsection{Evaluation Settings}
The replacement plans are compared when changing the parameters.
Only one parameter is changed, whereas the other parameters are fixed.
Five hours are allowed to operate our system, and the calculation is executed on the business day immediately preceding the delivery date.
The experimental settings are summarized in Table~\ref{tab:exp_result_edited}.
Numerical experiments were conducted under four different settings, as summarized in Table~\ref{tab:exp_ISG_result_edited}.
Note that when the hyperparameter $q_{\textrm{high}}$ was set to 0.5, we only focused on whether the average value of the distribution was above the threshold of the gas rate to determine a high-risk customer based on the definition of a risk function.

\subsection{Results and Discussion}
We use the same evaluation metrics shown in Section~\ref{sec:metric}.
The results of the evaluation metrics of the experiments are summarized in Table~\ref{tab:exp_result_edited}, enabling us to consider the following:

\renewcommand{\arraystretch}{1.0}
\begin{table}[H]
    \centering
    \begin{tabular}{c|c|c|c|c|c}
        Name & Exp 1 & Exp 2 & Exp 3 & Exp 4 & Exp 5 \\ \hline 
        $\alpha_{\textrm{high}}, q_{\textrm{high}}$ & 5 \%, 0.3 & 5 \%, \framebox[1.1\width]{0.5} & \framebox[1.1\width]{0 \%}, 0.3 & 5 \%, 0.3 & 5 \%, 0.3 \\
        $\alpha_{\textrm{mdr}}, q_{\textrm{mdr}}$ & 7 \%, 0.3 & 7 \%, \framebox[1.1\width]{0.5} & 7 \%, 0.3 & 7 \%, 0.3 & 7 \%, 0.3  \\
        $D_{lbd}$  & 2 & 2 & 2 & \framebox[1.1\width]{0} & 2   \\
        $D_{ph}$ & 3 & 3 & 3 & 3 & \framebox[1.1\width]{0}  \\[5 pt] \hline \hline
        Rate-average & 2.08 \% & -0.88 \% & -1.36 \% & 1.16 \%  & 0.49 \%  \\
        Rate-median & 4.10 \% & 2.57 \% & 0.09 \% & 3.87 \%  & 2.42 \% \\
        Success & 53.3 \% & 50.1 \% & 47.1 \% & 52.5 \% &  47.7 \%  \\
        Fail-out & 36.6 \% & 41.1 \% & 45.1 \% & 36.9 \% & 38.9 \%  \\
        \textbf{Fail-over(↓)} & 10.1 \% & 8.7 \% & \textbf{7.8 \%} & 10.6  \% & 13.4 \%  \\
        Fail-time(↓) & \textbf{0} & 1 & \textbf{0} & 1  & \textbf{0}  \\ 
        Time/customer(↓) & 13m06s & 13m00s & 12m05s & 13m38s & 11m24s  \\
        Distance/customer(↓) & 1.8 km & 1.8 km & 1.5 km & 2.0 km  & 1.2 km  \\
        \textbf{Run-out(↓)} & 28.1 & 31.6 & \textbf{21.6} & 26.6 & 29.6 \\
    \end{tabular}
    \caption{Hyperparameter settings for each experiment setting and evaluation metric of the experiment. 
    The surrounding rectangles highlight the differences in the settings between Exp 1 and the others. The best results are presented in bold.}
    \label{tab:exp_result_edited}
\end{table}
\renewcommand{\arraystretch}{1.0}

\begin{enumerate}
    \item ($\boldsymbol{q_{\textrm{high}}, q_{\textrm{mdr}}}$) The difference between Exp 1 and Exp 2 is the threshold for determining high- and moderate-risk customers.
    The experiments were conducted to observe the effect of considering the risk function defined in Section~\ref{sec:risk_func}.
    \begin{itemize}
        \item Exp 1 succeeded in increasing the success rate and decreasing the fail-out rate, but it failed to increase the fail-over rate.
        Because $q_{\textrm{high}}$ and $q_{\textrm{mdr}}$ were smaller in Exp 1, customers were more likely to be included in the customer list for replacement than in Exp 2.
        A smaller fail-out rate was observed by decreasing the parameters $q_{\textrm{high}}$ and $q_{\textrm{mdr}}$, which shows the effectiveness of considering the risk function described in Section~\ref{sec:risk_func}.
    \end{itemize}
    \item ($\boldsymbol{\alpha_{\textrm{high}}}$) In Exp 3, the remaining gas rate was set to 0\%, whereas it was set to as 5\% in Exp 1.
    \begin{itemize}
        \item Fewer run-out and fail-over events were observed in Exp 3 compared with Exp 1.
        Because $\alpha_{\textrm{high}}$ was smaller in Exp 3, it was possible to focus more on delivering to a customer who was about to experience a run-out event than with Exp 1.
        In other words, Exp 1 was more likely to achieve visit to customers whose remaining gas was not below 0\%, even if they were high-risk customers.
        The results indicate that we can focus on delivering to customers with a specific gas rate by changing $\alpha_{\textrm{high}}$.
    \end{itemize}
    \item (\textbf{$\boldsymbol{D_{lbd}}$}) Exp 4 did not force the addition of more moderate-risk customers without considering $D_{lbd}$, aiming to reveal the effectiveness of such dates.
    \begin{itemize}
        \item $D_{lbd}$ makes the number of moderate-risk customers increased.
        Notably, some customers were in the moderate-risk group in Exp 1. 
        However, they were not moderate-risk customers in Exp 4.
        These customers enabled us to obtain a smaller delivery area than in Exp 4.
        Such customers even had a low risk of experiencing a gas shortage.
        Therefore, more run-out events in Exp 1 were reported, with a larger number of visits and shorter distances/trips.
        However, for few high- and moderate-risk customers, we can obtain more candidates for visit to replace cylinders by setting $D_{lbd}$ larger than zero.
	This is a promising technique for realizing workload leveling.
        \item The two systems were competitive with respect to all evaluation metrics.
        This indicates that the influence of changing the trip division is slight when compared with the other parameters.
    \end{itemize}
    \item (\textbf{$\boldsymbol{D_{ph}}$}) The planning horizon was set to three and zero, in Exp 1 and Exp 5, respectively.
    \begin{itemize}
        \item In Exp 5, distance/customer was shorter than in Exp 1 because minimizing the delivery area for multiple days was not implemented when formulating the customer list for replacement. As a result, although problems to acquire customer lists in both Exp 1 and Exp 5 provide optimal solutions, the customer list of the first day in Exp 5 is equal to or smaller than that in Exp 1. Therefore, Exp 1 recorded worse results than Exp 5 regarding time/customer and distance/customer.
    \end{itemize}
    In conclusion, Exp 3 achieved the best results in fail-over and run-out events and had the best hyperparameter settings between Exp 1 and Exp 5.
\end{enumerate}

\section{Ablation Study on the Formulation to Acquire Customer List \label{sec:exp_ablation}}
The formulations of the proposed methods and when the depot need not be included in the rectangle, such as the formulation in~\cite{baller2019dynamic}, were compared.
The hyperparameter settings followed by Exp 3 are described in Appendix~\ref{sec:appendix_dldr}.
\begin{wrapfigure}[18]{r}[3mm]{0.7\linewidth}
    \centering
    \includegraphics[width=\linewidth]{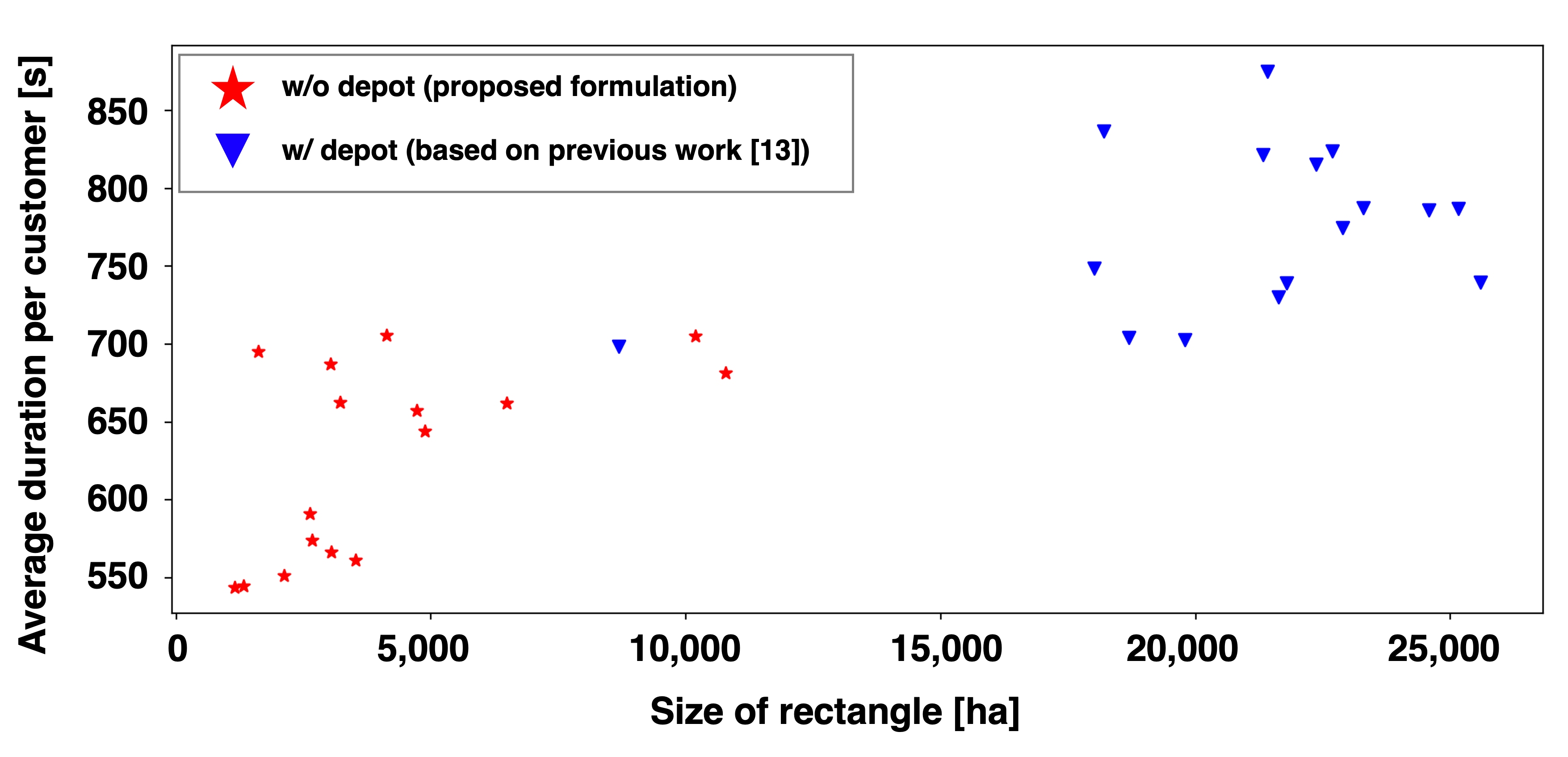}
    \caption{Scatter plot of the size of rectangle obtained by solving Problem~\ref{prob:multiBB} and average duration of delivery per customer. Blue triangle markers represent the result from an addition constraint in which the depot must be covered by a rectangle.}
    \label{fig:ablation}
\end{wrapfigure}
Figure~\ref{fig:ablation} shows the pairing of the size of the rectangle and the average delivery time, and each dot represents a trip.
With both methods, the size of the rectangle represented the cost of geographical delivery costs, and in fact, there was a correlation with respect to the average duration per customer.

From Figure~\ref{fig:ablation}, we can see that the average duration per customer for the proposed system (without depot) was smaller than that of the method in \cite{baller2019dynamic} (with depot).
This indicates that the proposed method acquires a more efficient delivery route.
This is because the depot is far from the center of gravity of the customers.
When a customer far from the depot must be delivered to, the rectangle covering the depot and the customer has already covered most of the other customers.

\end{document}